\documentclass[%
 reprint,
 amsmath,amssymb,
floatfix,
superscriptaddress
]{revtex4-2}
\usepackage{mathtools}
\usepackage{graphicx}
\usepackage{dcolumn}
\usepackage{bm}

\usepackage[caption=false]{subfig}

\usepackage{algorithm}
\usepackage[noend]{algpseudocode}
\algrenewcommand\algorithmicdo{}

\usepackage{braket}
\usepackage{amsmath}

\makeatletter
\renewcommand{\ALG@name}{Procedure}
\makeatother

\newcommand{\reffig}[1]{Fig.~\ref{#1}}

\newcommand{\refsec}[1]{Sec.~\ref{#1}}

\usepackage[linktocpage=true,
  colorlinks=true, 
  pdfborder={0 0 0},
  linkcolor=blue,
  citecolor=red,
  filecolor=yellow,
  urlcolor=blue,
  bookmarks,
  pdfauthor={},
]{hyperref}
\usepackage{microtype}
\usepackage{orcidlink}

\usepackage{ifthen}
\newcounter{is_qcircuit_used}
\usepackage[justification=raggedright,singlelinecheck=false]{caption}

\begin{document}

\preprint{APS/123-QED}

\title{
Accelerated spin-adapted ground-state preparation with nonvariational quantum algorithms
}
\author{Takumi Kobori\orcidlink{0009-0002-7940-7593}}
\email{takumi.kobori@phys.s.u-tokyo.ac.jp}
\affiliation{
Department of Physics, 
The University of Tokyo, 
Tokyo 113-0033, 
Japan
}

\author{Taichi Kosugi\orcidlink{0000-0003-3379-3361}}
\affiliation{
Quemix Inc.,
Taiyo Life Nihombashi Building,
2-11-2,
Nihombashi Chuo-ku, 
Tokyo 103-0027,
Japan
}

\affiliation{
Department of Physics,
The University of Tokyo,
Tokyo 113-0033,
Japan
}

\author{Hirofumi Nishi\orcidlink{0000-0001-5155-6605}}

\affiliation{
Quemix Inc.,
Taiyo Life Nihombashi Building,
2-11-2,
Nihombashi Chuo-ku, 
Tokyo 103-0027,
Japan
}

\affiliation{
Department of Physics,
The University of Tokyo,
Tokyo 113-0033,
Japan
}

\author{Synge Todo\orcidlink{0000-0001-9338-0548}}
\affiliation{
Department of Physics, 
The University of Tokyo, 
Tokyo 113-0033, 
Japan
}
\affiliation{
Institute for Physics of Intelligence, 
The University of Tokyo, 
Tokyo 113-0033, 
Japan}
\affiliation{Institute for Solid State Physics, 
The University of Tokyo, 
Kashiwa 277-8581, 
Japan}

\author{Yu-ichiro Matsushita\orcidlink{0000-0002-9254-5918}}
\affiliation{
Department of Physics,
The University of Tokyo,
Tokyo 113-0033,
Japan
}

\affiliation{
Quemix Inc.,
Taiyo Life Nihombashi Building,
2-11-2,
Nihombashi Chuo-ku, 
Tokyo 103-0027,
Japan
}

\affiliation{Quantum Materials and Applications Research Center,
National Institutes for Quantum Science and Technology (QST),
2-12-1 Ookayama, Meguro-ku, Tokyo 152-8550, Japan
}

\date{\today}

\begin{abstract}
Various methods have been explored to prepare the spin-adapted ground state—the lowest energy state within the Hilbert space constrained by externally specified values of the total spin magnitude and the spin-$z$ component.
In such problem settings, nonvariational methods incorporate penalty terms into the original Hamiltonian to enforce the desired constraints.
Such additional terms would require a quartic number of gates in terms of the number $n_{\textrm{spin}}$ of spins when naively implemented. To circumvent this computationally intensive scaling, this paper proposes a new nonvariational procedure to obtain the spin-adapted ground states that requires only quadratic scaling.
The proposed method consists of two stages: the first stage is to prepare a spin-magnitude adapted state and the second stage is post-processing for the desired $S_z$.
By separating into two stages, the procedure achieves the desired spin-adapted ground state while reducing the number of penalty terms from $O(n_{\textrm{spin}}^4)$ to $O(n_{\textrm{spin}}^2)$.
We conducted numerical experiments for spin-1/2 Heisenberg ring models and manganese trimer systems. 
The results confirmed the effectiveness of our method, demonstrating a significant reduction in gate complexity and validating its practical usefulness.
\end{abstract}

\maketitle 

\section{Introduction}
Exploration of complex spin systems plays a crucial role in deepening our understanding of quantum many-body physics~\cite{sandvik2010computational}. Quantum computers offer a promising platform for efficiently simulating such systems by leveraging their ability to represent the exploration space using qubits proportional to the system size~\cite{feynman2018simulating,nielsen2010quantum,orus2019tensor, georgescu2014quantum}.
As a result, the development of quantum algorithms for these tasks has been an active area~\cite{ low2019hamiltonian,bauer2020quantum,cerezo2021variational,cao2019quantum,wu2024variational}.
Among these, various methods have been proposed for preparing the ground state of quantum spin systems on quantum computers~\cite{childs2018toward,yoshioka2024hunting,lyu2023symmetry,farhi2000quantum,hejazi2024adiabatic,sun2021quantum,nishi2023optimal,kosugi2022imaginary,kadowaki1998quantum,mitarai2018quantum, motta2020determining}. 
In particular, considerable attention has been directed toward preparing spin-adapted ground states, which correspond to the lowest-energy eigenstates within Hilbert subspaces constrained to fixed values of the total spin magnitude and spin-$z$ component.
The explorations of such spin-restricted Hilbert subspaces are crucial for obtaining a more detailed and accurate understanding of problems in condensed matter physics and quantum chemistry~\cite{keller2016spin,sharma2012spin,pauncz2012spin,Planelles1994,gandon2024quantumcomputingspinadaptedrepresentations}.

Methods for calculating the spin-adapted ground state can generally be classified into variational and nonvariational approaches where penalty terms are commonly introduced into the original Hamiltonian to enforce spin symmetries~\cite{sugisaki2022adiabatic, seki2020symmetry,kuroiwa2021penalty,carbone2022quantum,lacroix2023symmetry,lyu2023symmetry,nielsen2010quantum,yoshioka2024hunting}.
In variational approaches, the penalty terms for the total spin magnitude are typically proportional to $(\langle\hat{S}^2\rangle -s^*(s^*+1))^2$, where $s^{*}$ is the target value for total spin magnitude, and are generally more computationally demanding than those for spin-$z$. 
Although the expression involves a fourth-order polynomial in $\hat{S}$, the number of circuit variants required scales as $O(n_{\textrm{spin}}^2)$, even under the worst-case encoding for a system of $n_{\textrm{spin}}$ spins to compute the expectation value of $\hat{S}^2$ from the measurement outcomes. 
While variational approaches may appear efficient at first glance, their practical performance is affected by additional factors, including the cost of classical optimization and the overhead associated with repeated executions of quantum circuits.

On the other hand, nonvariational approaches for calculating the spin-adapted ground state naively require $O(n_{\textrm{spin}}^4)$ quantum gates from a fourth-order polynomial of $\hat{S}$ and are therefore expected to be more resource-intensive~\cite{kosugi2022imaginary,xie2024probabilistic,liu2021probabilistic,nishi2024quadratic,farhi2000quantum,sugisaki2022adiabatic,nishiya2024first}.
Two straightforward methods are commonly employed in nonvariational approaches: preparing spin-adapted initial states and adding penalty terms to the Hamiltonian~\cite{sugisaki2022adiabatic,siwach2021filtering,stetcu2023projection,lacroix2020symmetry}.
In the first approach, if an initial state satisfying the desired values of $\hat{S}^2$ and $\hat{S}_z$ is prepared, the time evolution under a nonvariational method that preserves these symmetries can ideally lead to the desired spin-adapted state.
Although this approach is conceptually ideal, practical implementations face significant challenges. Various sources of errors, such as quantum hardware imperfections and algorithmic approximations (e.g., Suzuki-Trotter decompositions~\cite{childs2021theory,lloyd1996universal,jahnke2000error,ivan2008polynomial,trotter1959product,suzuki1976relationship} or Solovey-Kitaev decompositions~\cite{nielsen2010quantum,kitaev2002classical}) can cause deviations from the intended Hilbert subspace. 
These deviations may lead to the accumulation and amplification of errors during the computation. Therefore, in practical applications, it is crucial to enhance the energy alignment of the solution with the desired spin sector by introducing penalty terms. 
However, unlike variational approaches, naively introducing a penalty Hamiltonian involving spin operators in nonvariational methods is computationally inefficient. Representing penalty terms proportional to $(\hat{S}^2 -s^*(s^*+1))^2$ typically requires Hamiltonian-dependent unitary operations, such as $e^{-i\mathcal{H}t}$ in adiabatic time evolution, which in turn  demand $O(n_{\textrm{spin}}^4)$ gates.
Thus, accelerating the nonvariational approaches for obtaining spin-adapted ground states remains a key challenge and an important objective for the realization of fault-tolerant quantum computing in the future.

In this paper, we propose a novel procedure to reduce the computational cost of penalty terms in nonvariational methods for spin-rotationally symmetric Hamiltonians.
Our method consists of two main stages: the first involves preparing a spin-magnitude-adapted ground state, which is the lowest energy with the desired total spin magnitude, and the second performs post-processing to adjust the spin-$z$ component.
In the first stage, we introduce an extended penalty term to prepare the spin-magnitude-adapted ground state that adds an additional penalty term related to the spin-$z$ component to the penalty term associated with the total spin magnitude.
This modification allows us to eliminate the squared form of the original penalty term and reduce the number of terms from $O(n_{\textrm{spin}}^4)$ to $O(n_{\textrm{spin}}^2)$.
In the second stage, we introduce a post-processing method to obtain the desired $S_z$ states from the spin-magnitude-adapted state. The post-processing can be performed using a quantum circuit with a depth scaling of $O(\log^2 n)$. 
Typically, in physical systems, the terms of the Hamiltonian are smaller than $O(n_{\textrm{spin}}^2)$, which means that the square-root reduction in complexity for gate executions achieved by our method offers significant speedup.
To demonstrate the effectiveness of our proposed method, we performed numerical simulations on the Heisenberg ring model and the manganese trimer using adiabatic time evolution and probabilistic imaginary time evolution. The successful results demonstrate the versatility of our method.

\section{Background}
\subsection{nonvariational approaches to prepare ground states}
Currently, various studies are actively underway worldwide to efficiently calculate ground states on quantum computers. In this study, we have adopted and compared two ground state calculation methods based on nonvariational methods: imaginary time evolution (ITE) method and adiabatic time evolution (ATE) method.
This section briefly introduce these methods and discuss the challenges in calculating the spin-adapted ground states using nonvariational methods.
\subsubsection{ATE}
ATE is a nonvariational method that prepares the ground state of a target Hamiltonian by slowly interpolating from an initial Hamiltonian~\cite{farhi2000quantum,sugisaki2022adiabatic,nishiya2024first}.
To prepare the ground state of a Hamiltonian $\mathcal{H}_{\text{problem}}$ using ATE method, the procedure begins by preparing the ground state $\ket{\psi(0)}$ of an initial Hamiltonian $\mathcal{H}_{\text{initial}}$. The system is then evolved under a time-dependent Hamiltonian of the form $\mathcal{H}(t) = \left(1 - \frac{t}{T}\right) \mathcal{H}_{\text{initial}} + \frac{t}{T} \mathcal{H}_{\text{problem}}$ from $t=0$ to $t=T$, such that the adiabatic condition is satisfied. The evolution is typically implemented using discrete time steps, applying unitary operators of the form $e^{-i\mathcal{H}(t)dt}\ket{\psi(t)}$. The quantum circuit of ATE is illustrated in \reffig{fig:non_var_circuit}(a).
The procedure of ATE is as follows.
\begin{algorithm}[H]
\caption{Adiabatic Time Evolution (ATE)}
\begin{algorithmic}[1]
\Require Problem Hamiltonian $\mathcal{H}_{\text{problem}}$, Initial Hamiltonian $\mathcal{H}_{\text{initial}}$, Total evolution time $T$, Total time step $n$, Time schedule $\Delta t_i$
\Ensure Ground state of $\mathcal{H}_{\text{problem}}$
\State Initialize the state vector $\ket{\psi}$ as the ground state of $\mathcal{H}_{\text{initial}}$
\State $t \gets 0$
\For {$i = 0$ to $n$}
  \State $\mathcal{H}_t \gets \left(1 - \frac{t}{T}\right) \mathcal{H}_{\text{initial}} + \frac{t}{T} \mathcal{H}_{\text{problem}}$
  \State $\ket{\psi} \gets e^{-i\mathcal{H}_t\Delta t_{i}}\ket{\psi}$
  \State $t \gets t+\Delta t_{i}$
\EndFor
\State \Return $\ket{\psi}$
\end{algorithmic}
\end{algorithm}
Several factors must be taken into account when implementing ATE. 
First, ATE is computationally demanding because it requires a large total evolution time $T$ and sufficiently small time steps $\Delta t_i$, which are governed by the energy gap between the ground state and the first excited state. Consequently, selecting an appropriate initial Hamiltonian $\mathcal{H}_{\text{initial}}$ is crucial for computational efficiency.
Second, there exists a systematic error associated with the implementation of the time evolution operator $e^{-i\mathcal{H}(t)\Delta t}$ on a quantum circuit. This operation is typically achieved using the Suzuki-Trotter decomposition, which introduces decomposition errors.
\subsubsection{ITE and probabilistic ITE (PITE)}
ITE is a nonvariational method that applies the imaginary-time evolution of a Hamiltonian to obtain the ground state.
The state $\ket{\psi(\tau)}$ after evolving an initial state $\ket{\psi(0)}$ for an imaginary time $\tau$ is given by
\begin{align}
&\ket{\psi(\tau)} \propto e^{-\mathcal{H}\tau}\ket{\psi(0)}\notag \\
&\propto  \ket{\psi_0}\braket{\psi_0|\psi(0)}+\sum_{i>0} e^{-(E_i-E_0)\tau}\ket{\psi_i}\braket{\psi_i|\psi(0)} ,
\end{align}
where $\ket{\psi_i}$ and $E_i$ are the eigenstates and corresponding eigenvalues of the Hamiltonian $\mathcal{H}$, respectively.
Since higher-energy components are exponentially suppressed as $\tau$ increases, imaginary-time evolution effectively projects the state onto the ground state.
However, the imaginary time evolution operator $e^{-\mathcal{H}\tau}$ is non-unitary, and its implementation on a quantum computer is non-trivial. 
Several approaches to realize ITE on a quantum computer~\cite{jones2019variational,mcardle2019variational,yuan2019theory,motta2020determining,yeter2020practical,sun2021quantum,lin2021real,hejazi2024adiabatic} have been proposed. In this work, we focus on the probabilistic imaginary time evolution (PITE) method~\cite{kosugi2022imaginary,xie2024probabilistic,liu2021probabilistic,nishi2024quadratic}.
\begin{figure}[h]
  \centering
  \includegraphics[width=\linewidth]{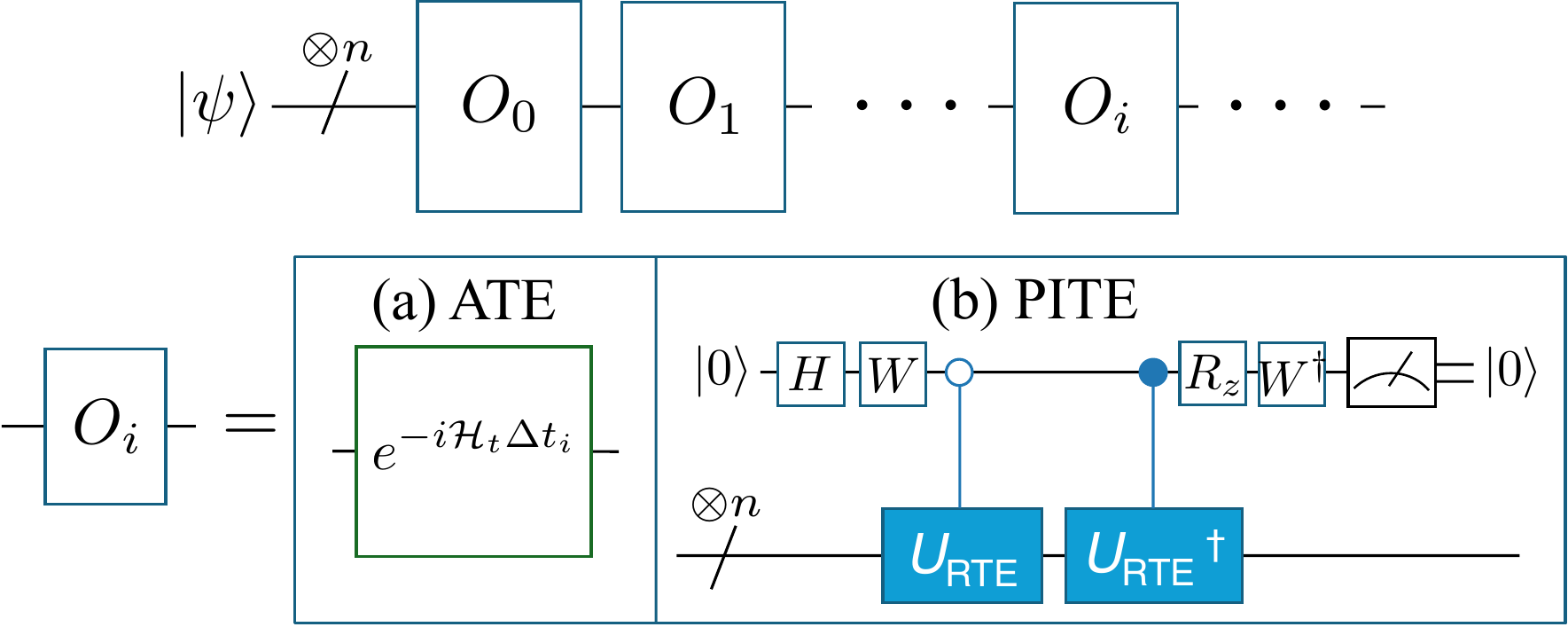}
  \caption{
  Circuits of the nonvariational approaches: (a) ATE and (b) PITE.
  In (b), $U_{\text{RTE}}=e^{-i\mathcal{H}s\Delta t}$.
  $n$ is the number of qubits used to represent the quantum system.
  }
  \label{fig:non_var_circuit}
  \end{figure}
PITE is a method to realize ITE on a quantum computer by using ancilla qubits, measurements, and post-selection. 
While several implementations of PITE have been proposed, in this work we adopted the method firstly proposed by Kosugi {\it et al.}~\cite{kosugi2022imaginary}, which approximately realizes ITE with only one additional ancilla qubit and controlled-$e^{-i\mathcal{H}t}$.
For implementing the non-unitary operator $\mathcal{M} = m_0 e^{-\mathcal{H}\tau}$, where $m_0$ is an adjustable constant satisfying $0<m_0<1$ and $m_0\neq 1/\sqrt{2}$, we extend the $n$-qubit system $\ket{\psi}$ to an $(n+1)$-qubit system by introducing an ancilla qubit and consider an unitary operation $U_{\mathcal{M}}$ defined as
\begin{align}
U_{\mathcal{M}} = \begin{pmatrix}
\mathcal{M} & \sqrt{1-\mathcal{M}^2} \\
-\sqrt{1-\mathcal{M}^2} & \mathcal{M}
\end{pmatrix}.
\end{align}
By using this unitary operation $U_{\mathcal{M}}$, the non-unitary operator $\mathcal{M}$ can be realized by initializing the ancilla qubit to $\ket{0}$, applying $U_{\mathcal{M}}$ and post-selecting the measurement outcome of the ancilla qubit to be $\ket{0}$.
$U_{\mathcal{M}}$ can be decomposed as 
\begin{align}\label{eq:PITEcircuit}
U_{\mathcal{M}} = ( W^{\dagger}\otimes I)\begin{pmatrix}
e^{i\kappa\Theta} & 0 \\
0 & e^{-i\kappa\Theta}
\end{pmatrix}(WH\otimes I),
\end{align}
where $H$ is the Hadamard gate, $W$ and $\Theta$ are described below,
\begin{align}
W &= \frac{1}{\sqrt{2}}\begin{pmatrix}
1 & -i \\
1 & i
\end{pmatrix}, \quad \Theta = \arccos\frac{\mathcal{M}+\sqrt{1-\mathcal{M}^2}}{\sqrt{2}},
\end{align} 
and $\kappa$ is a constant given by $\kappa = \mathrm{sgn}(m_0 -1/\sqrt{2})$.
The central component of $U_{\mathcal{M}}$ can be implemented using controlled unitaries $e^{\pm i\kappa\Theta}$~\cite{kosugi2022imaginary,nishi2024quadratic}.

Although the above methods are non-approximate, the computational cost of preparing controlled unitaries $e^{\pm i\kappa\Theta}$ is typically high.
In practice, these operations are approximated using a first-order expansion in the imaginary-time step $\Delta t$.
Expanding $\kappa\Theta$ as $\kappa\Theta = \theta_0 - \mathcal{H}s\Delta t + O(\Delta t^2)$, the unitary $U_{\mathcal M}$ in Eq.~(\ref{eq:PITEcircuit}) can be approximated as
\begin{align}
U_{\mathcal{M}} = (W^{\dagger}R_z(-2\theta_0)\otimes I)\notag\\ \begin{pmatrix}
e^{-i\mathcal{H}s\Delta t} & 0 \\
0 & e^{i\mathcal{H}s\Delta t}
\end{pmatrix}(WH\otimes I) + O(\Delta t^2),     
\end{align}
where $\theta_0=\kappa\arccos[(m_0+\sqrt{1-m_0^2})/\sqrt{2}]$, $s=m_0/\sqrt{1-m_0^2}$ and $R_z(\theta) = e^{-i\theta\sigma_z/2}$~\cite{kosugi2022imaginary,nishi2024quadratic}.
This approximation enables the implementation of imaginary-time evolution using controlled real-time evolution operators $e^{\pm i\mathcal{H}s\Delta t}$, as illustrated in Fig.~\ref{fig:non_var_circuit}(b).
The overall procedure of PITE is summarized as follows:
\begin{algorithm}[H]
  \caption{Probabilistic Imaginary Time Evolution (PITE)}
  \begin{algorithmic}[1]
  \Require Problem Hamiltonian $\mathcal{H}_{\text{problem}}$, Initial state $\ket{\psi}_0$, Total steps $N$, Time step $\Delta t$, Coefficient $m_0$
  \Ensure Ground state of $\mathcal{H}_{\text{problem}}$
  \State $s\gets m_0/\sqrt{1-m_0^2}$
  \State $\kappa\gets \text{sgn}(m_0 -1/\sqrt{2})$
  \State $\theta_0\gets \kappa\arccos\left[(m_0+\sqrt{1-m_0^2})/\sqrt{2}\right]$
  \State $\ket{\psi}\gets\ket{\psi}_0$
  \For {$t = 0$ to $N$}
    \State Add ancilla qubit $\ket{0}$ : 
    $\ket{\psi} \gets \ket{0}\otimes\ket{\psi}$
    \State Apply the circuit as shown in Fig.~\ref{fig:non_var_circuit}(b) : 
    \State \qquad $\ket{\psi}\gets( WH\otimes I)\ket{\psi}$
    \State \qquad $\ket{\psi}\gets(\ket{0}\bra{0}\otimes e^{-i\mathcal{H}s\Delta t} + \ket{1}\bra{1}\otimes I)\ket{\psi}$
    \State \qquad $\ket{\psi}\gets(\ket{0}\bra{0}\otimes I+ \ket{1}\bra{1}\otimes e^{i\mathcal{H}s\Delta t})\ket{\psi}$
    \State \qquad $\ket{\psi}\gets( W^{\dagger}R_z(-2\theta_0)\otimes I)\ket{\psi}$
    \State Measure the ancilla qubit and post-select to $\ket{0}$ : 
    \State \qquad $\ket{\psi} \gets (\bra{0}\otimes I)\ket{\psi}$
  \EndFor
  \State \Return $\ket{\psi}$
  \end{algorithmic}
\end{algorithm}
Like ATE, the time schedule of PITE can be adjustable, but in our numerical experiments, $\Delta t$ is set to a constant. 
Similarly, $m_0$ is typically fixed throughout the evolution. 
However, if the lowest eigenvalue of $\mathcal{H}_{\textrm{problem}}$ are known in advance, $m_0$ can be optimized accordingly to improve performance~\cite{nishi2023optimal,meister2022resource}.

Of course, there exist many nonvariational approaches for fault-tolerant quantum computing, such as standard eigenvalue solvers based on quantum phase estimation~\cite{nielsen2010quantum,yoshioka2024hunting}, as well as recent techniques including the quantum eigenvalue transformation of unitary matrices with real polynomials~\cite{dong2022ground} and the full quantum eigensolver~\cite{shijie2020full,wang2024improving}.
However, these methods are typically built with Hamiltonian-dependent operations. As discussed later, our research focuses on accelerating such Hamiltonian-dependent components and is not limited to only ATE and PITE. It is therefore applicable to a broad range of nonvariational quantum algorithms.

\subsection{Difficulty of nonvariational approach for spin-adapted ground states}
The simplest way to obtain the spin-adapted ground state is to add penalty terms of the spin operators to the original Hamiltonian.
However, since expectation values of physical observables cannot be measured during the execution of nonvariational algorithms, it is not possible to use the same penalty terms as in variational methods.
The penalty terms for nonvariational methods are based on the total spin magnitude $\hat{S}=\sum_{i=0}^{n_{\textrm{spin}}-1}\bm{\sigma}_{i}/2$ and the $z$-component spin operator $\hat{S}_z=\sum_{i=0}^{n_{\textrm{spin}}-1}\sigma_{iz}/2$ as
\begin{align}\label{ex:conventinal_penalty}
\mathcal{H}_{\text{penalty}} &= C_{S}(\hat{S}^2-s^*(s^*+1))^2 + C_z(\hat{S}_z-s_z^*)^2\notag\\
&\equiv C_{S} \mathcal{H}_{S} + C_{z} \mathcal{H}_{z},
\end{align}
where $n_{\textrm{spin}}$ is the number of spins, $\sigma_{i\mu}$ are the Pauli $\mu$ matrices acting on $i$-th spin ($\mu$ is $x$, $y$, or $z$), $C_S$ and $C_z$ are positive constants~\cite{carbone2022quantum}. The parameters $s^*$ and $s_z^*$ are the target total spin magnitude and $z$-component, respectively. By adding the penalty terms to the original Hamiltonian, the resulting ground state will satisfy the desired spin properties.

Here we discuss the number of terms involved in the penalty terms.
In particular, we compare the number of terms between $\mathcal{H}_{S}$ and $\mathcal{H}_{z}$.  $\mathcal{H}_{z}$ has fewer terms than $\mathcal{H}_{S}$ because of the squared form of $\hat{S}^2$ in $\mathcal{H}_{S}$.
To discuss the $\mathcal{H}_{S}$ penalty term in more detail, we consider the term $\bm{\sigma}_i\cdot\bm{\sigma}_j$.
If $i=j$, we have $\bm{\sigma}_i\cdot\bm{\sigma}_i = \bm{\sigma}_i^2 = 3I$.
For $i\neq j$,
\begin{align}
  \begin{split}
      \bm{\sigma}_i\cdot\bm{\sigma}_j
      \ket{00}_{ij}
      &=\ket{00}_{ij}
      \\
      \bm{\sigma}_i\cdot\bm{\sigma}_j
      \ket{01}_{ij}
      &=2\ket{10}_{ij}-\ket{01}_{ij}
      \\
      \bm{\sigma}_i\cdot\bm{\sigma}_j
      \ket{10}_{ij}
      &=2\ket{01}_{ij}-\ket{10}_{ij}
      \\
      \bm{\sigma}_i\cdot\bm{\sigma}_j
      \ket{11}_{ij}
      &=\ket{11}_{ij}.
      \\
    \end{split}
\end{align}
By using the swap matrices $U^{S}_{ij}$ that exchanges spin $i$ and spin $j$, the term $\bm{\sigma}_i\cdot\bm{\sigma}_j$ can be expressed as 
\begin{equation}
  \bm{\sigma}_i\cdot\bm{\sigma}_j = 2U^{S}_{ij} - I.
\end{equation} 
Using the above relation, the $\hat{S}^2$ term can be rewritten as
\begin{equation}
  \hat{S}^2 = \frac{n_{\textrm{spin}}(4-n_{\textrm{spin}})}{4} + \sum_{i > j}U^{S}_{ij}.
\end{equation}
It indicates that in the variational method, performing $O(n_{\textrm{spin}}^2)$ measurements is sufficient to evaluate the penalty term~\cite{Shirai2023}.
Moreover, depending on the encoding method, it is possible to evaluate multiple $\langle U^{S}_{ij}\rangle$ terms using a single circuit.
For example, in a naive encoding where each spin is assigned to a distinct qubit, the penalty term can be evaluated using only $O(n_{\mathrm{spin}})$ different circuits.

The $\mathcal{H}_{S}$ penalty term can be expressed by $U^{S}_{ij}$ as
\begin{align}
\mathcal{H}_{S} &= (\hat{S}^2-s^*(s^*+1))^2\notag \\
&= a^2+2a\sum_{i > j}U^{S}_{ij}+\left(\sum_{i >j}U^{S}_{i j}\right)^2,
\end{align}
where $a = n_{\textrm{spin}}(4-n_{\textrm{spin}})/4-s^*(s^*+1)$~\cite{lowdin1969exchange}.
The last term involves all pairwise products of swap operators, resulting in a total of $O(n_{\textrm{spin}}^4)$ terms. 
Consequently, the evaluation of the penalty Hamiltonian $\mathcal{H}_{S}$ in nonvariational methods incurs a worst-case computational complexity of $O(n_{\mathrm{spin}}^4)$, whereas variational methods require only $O(n_{\mathrm{spin}}^2)$ measurement terms~\cite{kattem2022variational, kuroiwa2021penalty}.
While nonvariational methods can achieve some degree of reduction by parallelizing the evaluation of the $U_{ij}^{S}$ terms, for example in the case of PITE, this can be achieved by increasing the number of ancilla qubits~\cite{Yi2025}, such parallelization can also be incorporated into variational methods. 
This underscores the need for the development of more fundamental approaches to reduce the computational cost in nonvariational approaches further.

\section{Method}
In this section, we propose a novel method to address the issue mentioned above in spin-adapted state preparation when the original system Hamiltonians have spin-rotational symmetry. Our approach enables a square-root reduction in the number of terms required for the penalty term.
Our proposed method is based on the following two main stages:
\begin{enumerate}
  \item Prepare the spin-magnitude-adapted ground state with maximal spin-$z$ component (i.e. $S_z=s^*$) using a nonvariational method.
  \item Apply a repeat-until-success post-processing with rotation operator within the subspace that conserves the total spin, followed by a projection onto the subspace corresponding to a desired Hamming weight of given $S_z$.
\end{enumerate}
By introducing the two-stage procedure as described above and obtaining the ground state for the Hamiltonian with the properly constructed penalty terms using nonvariational approaches, we successfully reduce the number of terms in the penalty Hamiltonian from $O(n_{\textrm{spin}}^4)$ to $O(n_{\textrm{spin}}^2)$, thereby achieving a speedup.
It is worth noting that the proposed method is expected to have broad applicability since many Hamiltonians exhibit spin-rotational symmetry~\cite{moudgalya2028entanglement, chilton2022molecular,essler2005one,rajca1994organic}.

\subsection{Prepare the spin-magnitude-adapted ground state with maximal $S_z$}\label{subsec:prepare}
By restricting the target ground state to the subspace of $S_z=s^*$, we can reduce the number of terms in the penalty Hamiltonian for the total spin magnitude to $O(n_{\textrm{spin}}^2)$.
The penalty term can be written as
\begin{align}
\mathcal{H}'_{\text{penalty}} &= C'_{S}(\hat{S}^2-s^*(s^*+1))-C'_{z}(\hat{S}_z-s^*)\\
&\equiv C'_{S}\mathcal{H}'_{S} + C'_{z}\mathcal{H}'_{z}
\end{align}
where $C'_{S}$ and $C'_{z}$ are positive constants.
By eliminating the quadratic terms, we successfully reduce the number of terms from $O(n_{\textrm{spin}}^4)$ to $O(n_{\textrm{spin}}^2)$. The eigenvalues of this modified penalty Hamiltonian for the spin eigenfunctions $\ket{s, s_z}$ are given by 
\begin{align}
  E_{\mathrm{penalty}} &(s, s_z)
    \equiv
        \langle s, s_z |
        (C'_S\mathcal{H}'_S + C'_z\mathcal{H}'_{z})
        | s, s_z \rangle
    \nonumber \\
    &=
        C'_S
        \left(
            s (s + 1)
            -
            s^*
            (s^* + 1)
        \right)
        -
        C'_{z}
        (s_z - s^*),
\end{align}
where $s$ and $s_z$ denote the total spin and its $z$-component quantum numbers, with $s$ taking integer or half-integer values depending on the spin system, and $s_z \in \{-s, -s+1, \ldots, s\}$.
The minimum eigenvalue with respect to $s_z$ for a given $s$ is obtained at $s_z=s$:
\begin{align}
  E_{\mathrm{penalty}} (s)
    &\equiv
        \min_{s_z} E_{\mathrm{penalty}} (s, s_z)
    \nonumber \\
    &=
        E_{\mathrm{penalty}} (s, s)
    \nonumber \\
    &=
            C'_S
            \left(
                s (s + 1)
                -
                s^*
                (s^* + 1)
            \right)
            -
            C'_{z} (s - s^*).
\end{align}
Since this is a quadratic function of $s$, to attain the minimum value at $s=s^*$, it must satisfy the inequality $
E_{\mathrm{penalty}} (s^* - 1)
>
E_{\mathrm{penalty}} (s^*)$ and $
E_{\mathrm{penalty}} (s^*)
<
E_{\mathrm{penalty}} (s^* + 1)
$.
Therefore, the condition for the minimum eigenvalue of the penalty Hamiltonian is given by
\begin{align}
  2 s^*
  <
      \frac{C'_{z}}{C'_S}
  <
      2 (s^* + 1).
  \label{spin_constraints_1q:cond_for_penalty_m2_z1}
\end{align}\label{eq:c_inequality}
The above discussion applies to an arbitrary state, including mixed states induced by quantum noise, because the subspaces labeled by $\ket{s,s_z}$ are orthogonal, allowing their contributions to $E_{\mathrm{penalty}}$ to be treated independently. 

Under this condition, the spin-adapted ground state with maximal $S_z$ can be efficiently prepared by nonvariational methods.
Also, the ground state with minimal $S_z$, i.e.,  $S_z=-s^*$, can be obtained by the similar procedure.
It should be noted that the coefficients $C'_S$ and $C'_z$ must be chosen sufficiently large to prevent contamination from other undesired spin states. Although the above discussion focuses solely on the penalty terms, the total Hamiltonian can be written as $\mathcal{H}_{\textrm{problem}}=\mathcal{H}_{\textrm{system}}+\mathcal{H}_{\textrm{penalty}}$, where $\mathcal{H}_{\textrm{system}}$ is the original system Hamiltonian. If the penalty coefficients are too small, the resulting ground state of the total Hamiltonian may not correspond to the desired total spin magnitude. 
Therefore, the coefficients must be set large enough so that the terms $\mathcal{H}_{\textrm{penalty}}$ are dominant. On the other hand, excessively large coefficients require a smaller time step $\Delta t$ in the nonvariational evolution, leading to a decrease in the computational efficiency.
Of course, the penalty strengths must be appropriately optimized for each specific system. However, if we focus solely on the energy gap of the penalty term $\mathcal{H}_{\textrm{penalty}}$, the optimal choice is given by $C'_{z}/C'_{s} = 2s^{*} + 1$, and the resulting energy gap of $\mathcal{H}_{\textrm{penalty}}$ will be $C'_{s}$.

\subsection{Post processing to achieve the desired $S_z$}\label{subsec:post}
Post processing lies in a projection operator to vary $S_z$ onto the desired $S_z$ state while maintaining the total spin magnitude.
Such a projection operation can be implemented using a rotation operator within the spin-$S$ subspace.
Here, we consider the rotation operator around the $y$-axis defined as $U_y(\theta) = e^{-i\theta\hat{S}_y}$, where $\hat{S}_y=\sum_{i} \sigma_{iy}/2$. Due to the additive structure of $\hat{S}_y$, the rotation operator can be implemented by a depth-1 circuit using only single-qubit rotations as $U_y(\theta) = e^{-i \theta \sigma_y/2 } \otimes \cdots \otimes e^{-i \theta \sigma_y/2 }$.
Importantly, due to the spin-rotational symmetry of the Hamiltonian, the rotated state remains within the desired ground state corresponding to the total spin magnitude $S=s^*$. This allows us to explore different $S_z$ states while ensuring that the system remains in the desired spin subspace.

\begin{figure}[tbp]
\centering
\includegraphics[width=\columnwidth]{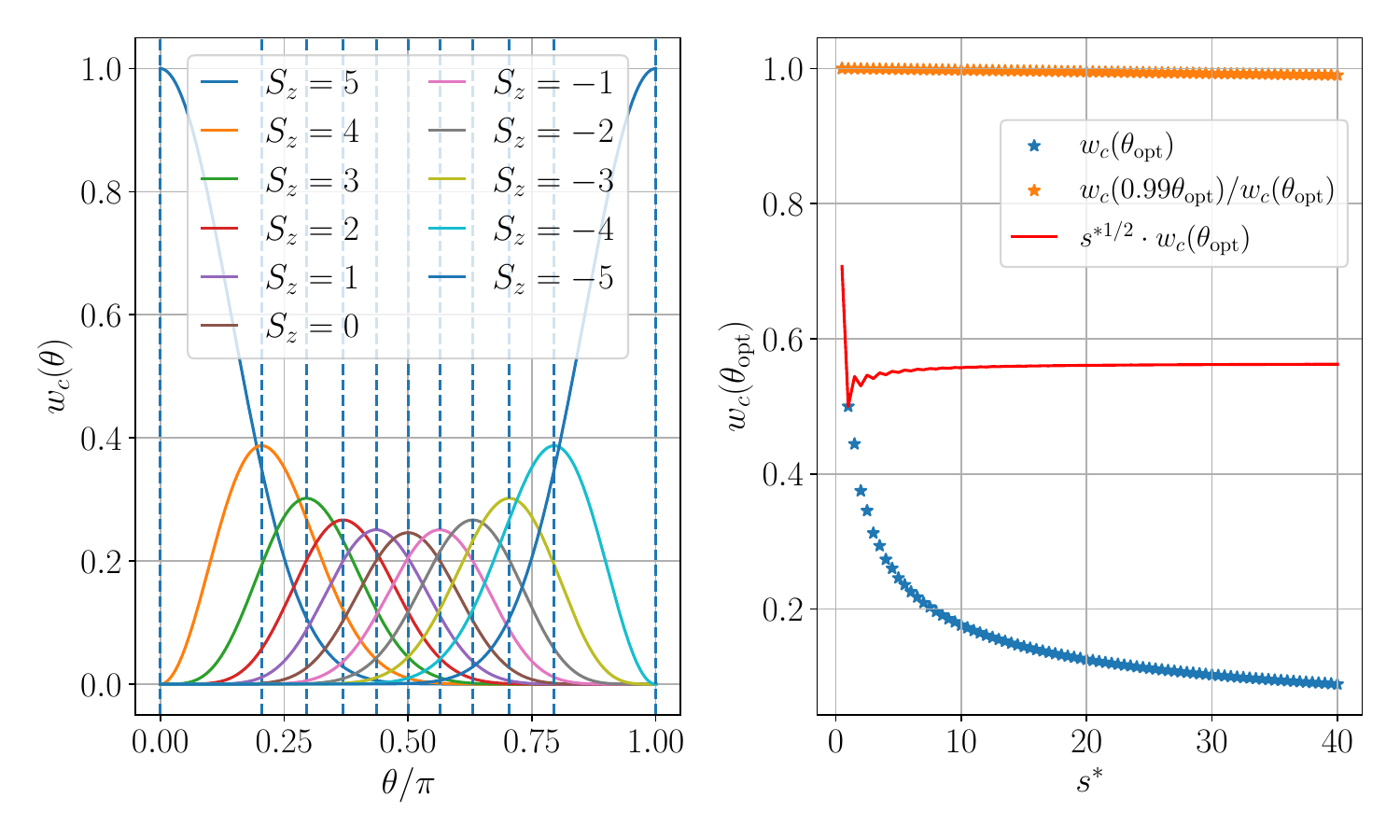}
\caption{(Left) $\theta$-dependence of the weight of the various $S_z$ states for $s^*=5$. Each line represents a different $S_z$ and each vertical dotted line represents $\theta_{\mathrm{opt}}$. (Right) Dependency of minimum $w_c(\theta_{\mathrm{opt}})$ in $S_z$ on $s^{*}$.
$S_z$ are set to $S_z=0$ or $S_z=1/2$.}
    \label{fig:weight}
\end{figure}

The remaining task is to determine the optimal rotation angle $\theta$ corresponding to the desired $s^*_z$.
For later projection onto the target $s_z^*$ eigenstate from the obtained state $\ket{s^*,s_z=s^*}$, it is necessary to maximize the probability of obtaining the desired $S_z$ state.
The weight of the target state in the rotated state is given as 
\begin{align}
  w_{\mathrm{c}} (\theta)
  \equiv
      \left|
      d^{s^*}_{s_{z}^* s^*} (\theta)
      \right|^2=\left|\langle s^*, s_z^* |
  e^{-i \theta \hat{S}_y}
  | s^*, s^* \rangle\right|^2,
\end{align}
where $d^{s^*}_{s_{z}^* s^*}$ is the Wigner d-matrix element~\cite{wigner2012group}.
It can be expressed as
\begin{align}
  d^{s^*}_{s_{z}^* s^*} (\theta)
  =
      \sqrt{
          \frac{
              (2s^*)!
          }{
            s^*_{+}!s^*_{-}!
          }
      }
      \left(
          \cos \frac{\theta}{2}
      \right)^{s^*_{+}}
      \left(
          \sin \frac{\theta}{2}
      \right)^{s^*_{-}}
  \label{spin_constraints_1q:Wigner_D_matrix}
\end{align} 
where $s^{*}_{\pm}=s^{*}\pm s^{*}_{z}$.
The optimal $\theta$ for the maximum $d_{s_z^* s^*}^{s^*}$ is obtained by setting $\partial d^{s^*}_{s_z^* s^*}(\theta)/\partial\theta = 0$,
which yields 
\begin{align}
  \theta_{\mathrm{opt}}
  =
      2 \arcsin \sqrt{\frac{s^* - s_z^*}{2 s^*}}.
\end{align}
The optimality of this value is also supported by the numerical results shown in Fig.~\ref{fig:weight}~(Left).
The weight of obtaining the desired $S_z$ state is given by
\begin{align}\label{eq:opt_weight}
  w_{\mathrm{c}} (\theta_{\mathrm{opt}})
  =
      \frac{
          (2 s^*)!
      }{
        s^*_{+}!s^*_{-}!
      }
      \frac{
          (s^*_{+})^{s^*_{+}}
          (s^*_{-})^{s^*_{-}}
      }{
        (2 s^*)^{2s^*}
      }.
\end{align}
\begin{figure}[tbp]
  \begin{center}
  \includegraphics[width=\linewidth]{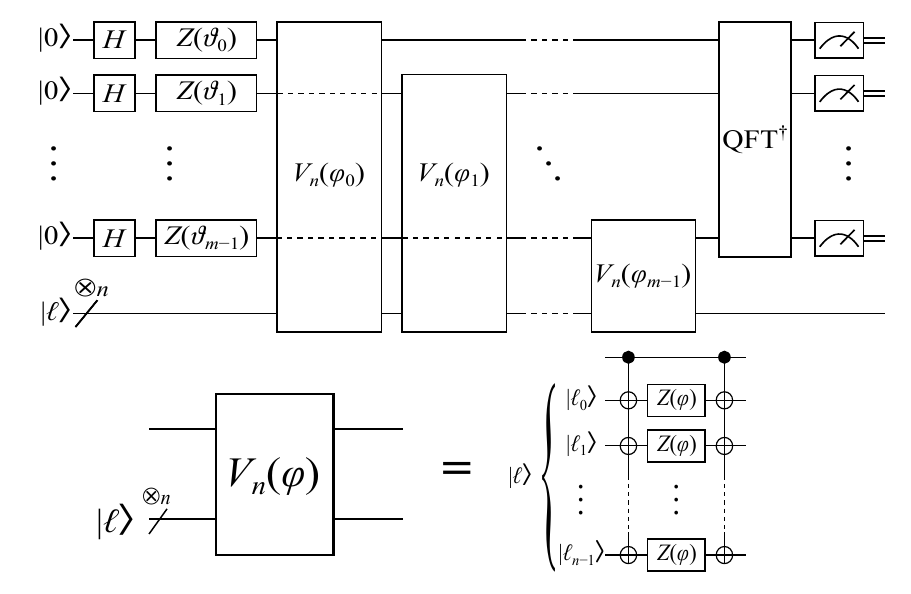}
  \end{center}
  \caption{
  Circuit for projecting an $n$-qubit input $\ket{l}$ onto the subspace with a specified Hamming weight, using $m = \lceil \log_{2} n \rceil$ ancilla qubits. The circuit parameters are given by $\vartheta_k = 2^{k-m}\pi n$ and $\varphi_k = -2^{k-m}\pi$~\cite{zi2024shallow}. Post-selection on the measurement outcomes of the ancilla qubits ensures the correct projection.
  }
  \label{fig:hamming}
\end{figure}

The second operation is the projection onto the desired $S_z$ state.
In general, the $S_z$ eigenstates can be described in terms of the computational basis states $\ket{j}$.
Specifically, if $n_1$ denotes the number of ones in the binary representation of $j$, the corresponding $S_z$ eigenstate is given by $\ket{j}$ where $n_1=n_{\textrm{spin}}/2-S_z$.
Thus, the projection operation can be realized by projection onto the subspace with Hamming weight $n_{\textrm{spin}}/2-S_z$, as illustrated in Fig.~\ref{fig:hamming}.
The depth of the circuit implementing this projection is $O(\log^2 n)$ with a gate set consisting of arbitrary single-qubit unitaries and CNOT gates~\cite{zi2024shallow}.
To analyze the success probability of post-selection, we plotted the dependency of the minimum $w_c(\theta_\mathrm{opt})$ across $S_{z}$ on given $s^{*}$, which characterizes the minimum success probability of post-selection with $\theta_{\textrm{opt}}$, in Fig.~\ref{fig:weight} (Right). The results indicate that the success probability scales as $O(s^{*-1/2})$. The behavior can also be derived from the properties of Catalan numbers~\cite{stanley2015catalan}.
Because the failure states can be reused, the post-processing constitutes a repeat-until-success procedure. By choosing an appropriate rotation angle, the expected number of repetitions scales as $O(s^{*1/2})$, implying that this procedure is typically more efficient than the first stage of our approach described in \refsec{subsec:prepare}.
Obtaining low-$s^{*}$ spin-adapted states is often of primary importance, implying that small $s^{*}$ values are sufficient in practice~\cite{vasiliev2018milestpnes}.
Thus, when exploring low-$s^{*}$ states, the present procedure is more efficient.

It should be noted that for binary encoding~\cite{Mootz2024}, as shown in Appendix~\ref{sec:binary} and used, for instance, in the calculation of the manganese trimer, a modification of the circuit is necessary. 
With binary encoding, a $d$-level system can be represented by $O(\log d)$ qubits. However, the depth of the unitary operation may increase exponentially in the worst case, as such encodings generally sacrifice the simple expressivity of quantum gates~\cite{sawaya2020resource}. Nonetheless, in condensed matter physics, $d$ of each spin or molecule is typically small, and the resulting overhead is not practically significant.

Regarding the imperfection in $\theta_{\mathrm{opt}}$, the algorithm is relatively robust because the landscape of $w_c(\theta)$ is flat in the vicinity of $\theta_{\mathrm{opt}}$, as shown in Fig.~\ref{fig:weight} (left).
To quantify this effect, we plot the ratio $w_c(0.99\,\theta_{\mathrm{opt}})/w_c(\theta_{\mathrm{opt}})$ in Fig.~\ref{fig:weight} (right). As the system size $S$ increases, the gap between $w_c(\theta_{\mathrm{opt}})$ and $w_c(0.99\,\theta_{\mathrm{opt}})$ becomes larger. This behavior originates from the fact that the landscape of $w_c(\theta)$ becomes sharper with increasing $s^{*}$.
Nevertheless, even for large $s^{*}$, the relative deviation remains small, indicating that the algorithm is robust against moderate inaccuracies in $\theta_{\mathrm{opt}}$. 

\subsection{Execution cost complexity comparisons}
To compare the time complexity of our method with previous approaches, particularly ATE and PITE, we focus solely on the penalty terms $\mathcal{H}_{\textrm{penalty}}$ and $\mathcal{H'}_{\textrm{penalty}}$ for simplicity.
The precise algorithmic cost generally depends on various factors, including the energy spectrum of the system Hamiltonian and the choice of initial states.
In the present discussion, we therefore concentrate on the energy gaps between the ground state and the first excited state induced by the penalty Hamiltonians.
They are evaluated as follows: $\Delta_{\textrm{quartic}}=\min(4C_S s^{*2},C_z)$, and $\Delta_{\textrm{ours}}=\min (2C'_{S}(s^*+1)-C'_{z}, -2C'_{S}s^*+C'_{z}))$ for $\mathcal{H}_{\textrm{penalty}}$ and $\mathcal{H'}_{\textrm{penalty}}$, respectively.
While the coefficients $C^{(')}_S$ and $C^{(')}_z$ should in general be optimized with respect to the system Hamiltonian, we here focus on a simplified scenario in which the optimization is carried out only in terms of the penalty gaps $\Delta_{\textrm{quartic}}$ and $\Delta_{\textrm{ours}}$.
This assumption is justified, for instance, when no prior information about the system Hamiltonian is available.
Under this condition, we set $C_z / C_S = O(s^{*2})$ and $C'_z / C'_S = 2 s^* + 1$, leading to $\Delta_{\textrm{quartic}} = O(C_S s^{*2})$ and $\Delta_{\textrm{ours}} = C'_S$.
Since this difference between $C_{S}$ and $C'_{S}$ complicates a direct comparison, we tune the constant factors such that the penalty gaps satisfy $\Delta_{\textrm{quartic}} \simeq \Delta_{\textrm{ours}}$ by setting $C'_S = C_S s^{*2}$.
Throughout the following discussion, the coefficients $C^{(')}_S$ and $C^{(')}_z$ are fixed accordingly.

The time complexity of ATE and PITE can be evaluated as the product of the cost of executing each subroutine $O_i$ and the total number of subroutines, as illustrated in \reffig{fig:non_var_circuit}.
The former is evaluated by the implementation cost of $U_{\mathrm{RTE}}$ or controlled-$U_{\mathrm{RTE}}$ and the success probability of each subroutines $O_i$. 
There are several approaches to implementing these operators. 
Here, we focus on two well-established methods: product-formula-based simulation and approaches based on quantum signal processing and quantum singular value transformation (QSP–QSVT)~\cite{childs2018toward,dalzell2025quantum,low2019hamiltonian,gilyen2019quantum}.
We consider a time-independent Hamiltonian $\mathcal{H}$ acting on an $n$-qubit system, written as a sum of $L$ Pauli operators, $\mathcal{H} = \sum_{j=1}^{L} h_j P_j$, where $h_j \in \mathbb{R}$ and $P_j$ are Pauli strings, and we define the $1$-norm of the Hamiltonian as $\|\mathcal{H}\|_1 = \sum_{j=1}^{L} |h_j|$.
Using a $k$th-order product formula, the gate complexity for implementing $e^{-i C \mathcal{H}}$ and its controlled version scales as
\begin{equation}\label{eq:product}
O\!\left(n L (C \|\mathcal{H}\|_1)^{1+\frac{1}{k}}\right),    
\end{equation}
where $C$ is a constant prefactor~\cite{dalzell2025quantum}.
In contrast, QSP–QSVT-based Hamiltonian simulation achieves a gate complexity scaling as
\begin{equation}\label{eq:qsvt}
O\!\left(n L\, C \|\mathcal{H}\|_1\right),    
\end{equation}
at the worst case~\cite{dalzell2025quantum,low2019hamiltonian,gilyen2019quantum}.
Depending on the structure of the Hamiltonian, these costs may be reduced, for example, by parallel gate execution.
In terms of the implementation cost of the controlled-$U_{\mathrm{RTE}}$, the cost increases by a constant factor, meaning that it remains on the same scale as the cost of implementing $U_{\mathrm{RTE}}$~\cite{barenco1995elementary,william2025halving,nielsen2010quantum}.
It should be noted that the QSP-QSVT-based implementation requires additional ancilla qubits, whereas the product-formula-based implementation does not require ancilla qubits~\cite{childs2018toward, dalzell2025quantum,gilyen2019quantum}.
Regarding the success probability, ATE is not probabilistic process and depends only on the initial state for PITE~\cite{kosugi2023exhaustive}, rather than on the energy spectrum. Therefore, we do not consider it further here.

The number of subroutines $n_{\textrm{step}}$ is determined by the energy gap $\Delta$ between the ground state and the first excited state.
For ATE, the number of steps scales as $n_{\textrm{step}} = O(1/\Delta_{\textrm{min}}^2)$~\cite{albash2018adiabatic} where $\Delta_{\textrm{min}}$ is the minimum gap over the entire adiabatic evolution, while for PITE it scales as $n_{\textrm{step}} = O(1/\Delta)$~\cite{kosugi2023exhaustive}.
Under the assumption $\Delta_{\textrm{quartic}} \simeq \Delta_{\textrm{ours}}$, achieved by the appropriate choice of $C^{(')}_S$ and $C^{(')}_z$, the dependence on $\Delta$ becomes identical for the two penalty constructions.

Based on the above discussion, the ratios of the time complexity of our method to that of the previous method are summarized as follows:
\begin{equation}
    O((s^{*2}/n_{\textrm{spin}}^2)^{1+\frac{1}{k}} / n_{\textrm{spin}}^{2})
\end{equation}    
for the product-formula-based implementation, and
\begin{equation}
    O(s^{*2} / n_{\textrm{spin}}^{4})
\end{equation}
for the QSP-QSVT-based implementation.
These scalings are derived from $L=O(n^2_{\textrm{spin}})$ and $||\mathcal{H}||_{1}=O(n^2_{\textrm{spin}})$ for $\mathcal{H}'_{\textrm{penalty}}$ and $L=O(n^4_{\textrm{spin}})$ and $||\mathcal{H}||_{1}=O(n^4_{\textrm{spin}})$ for $\mathcal{H}_{\textrm{penalty}}$.
According to these scalings, our algorithm can realize a certain degree of acceleration.
On the other hand, we must take into account the fact that Eqs.~(\ref{eq:product}, \ref{eq:qsvt}) represent worst-case estimates.
In particular, in the QSP–QSVT method, the factor $nL$ originates from the construction cost of the block-encoding and can be reduced by exploiting the structure of the given Hamiltonian~\cite{dalzell2025quantum}. If we assumed the factor $nL$ could be omitted, Eq.~(\ref{eq:qsvt}) would reduce to $O(C||\mathcal{H}||_1)$, and the corresponding ratio would be $O(s^{*2} / n_{\textrm{spin}}^{2})$.
In this case, the total cost may exceed that of the previous method on large $s^*$.
Moreover, in practical settings, the efficiency of the proposed method strongly depends on the system Hamiltonian and the choice of the initial state.
Consequently, the actual cost may differ from the above discussion. Nevertheless, our method can still provide substantial acceleration, since reducing the number of terms in the penalty Hamiltonian reduces the construction cost of $U_{\textrm{RTE}}$ and controlled-$U_{\textrm{RTE}}$.
\begin{figure*}[tbp]
    \centering
    \includegraphics[width=\linewidth]{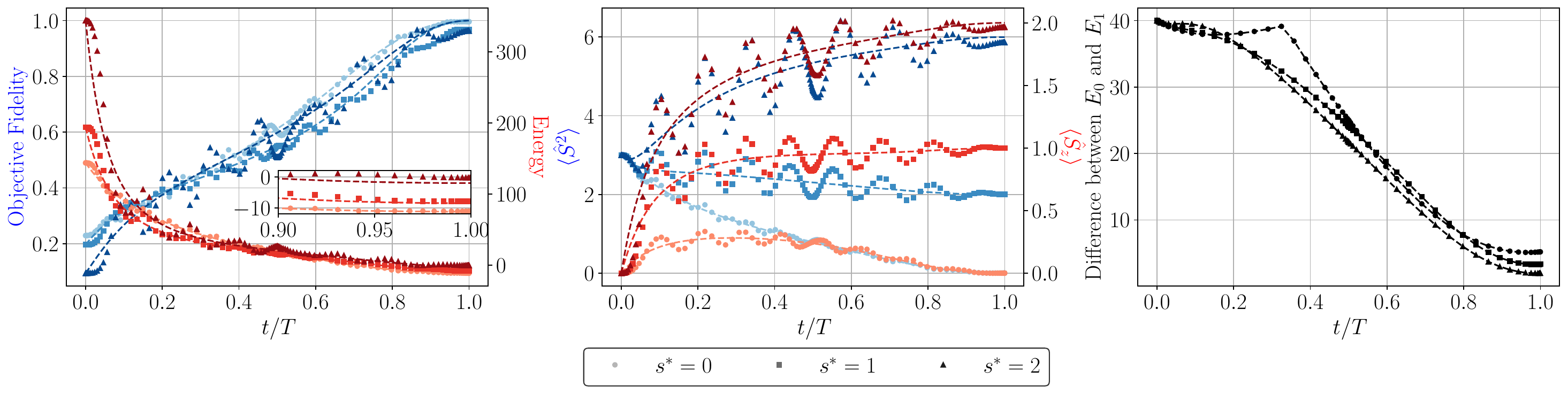}
    \caption{Results of our proposed methods for preparing spin-adapted ground states of Heisenberg ring models by ATE. 
    Different markers represent the target spin quantum numbers $s^*$: circles, squares, and rectangles correspond to $s^*=0$, $s^*=1$, and $s^*=2$, respectively.
The vertical axes represent various physical observables.
``Objective Fidelity"~(blue) denotes the fidelity between the target state and the instantaneous state.
``Energy"~(red), ``$\langle \hat{S}^2\rangle$"~(blue), and ``$\langle \hat{S}_z\rangle$"~(red) represent the expectation values of $\mathcal{H}_{\mathrm{problem}}$, $\hat{S}^2$, and $\hat{S}_z$, respectively.
The dotted lines in each panel represent the exact values of the corresponding observables, obtained by exact diagonalization of the instantaneous Hamiltonian $\mathcal{H}_t$.
The right panel shows the instantaneous energy gap between the ground state and the first excited state of the instantaneous Hamiltonian $\mathcal{H}_{t}$. Each point is computed by exact diagonalization of $\mathcal{H}_{t}$.
}
    \label{fig:result_heisenberg_adiabatic}
\end{figure*}
\begin{figure}[tbp]
\centering
\includegraphics[width=\columnwidth]{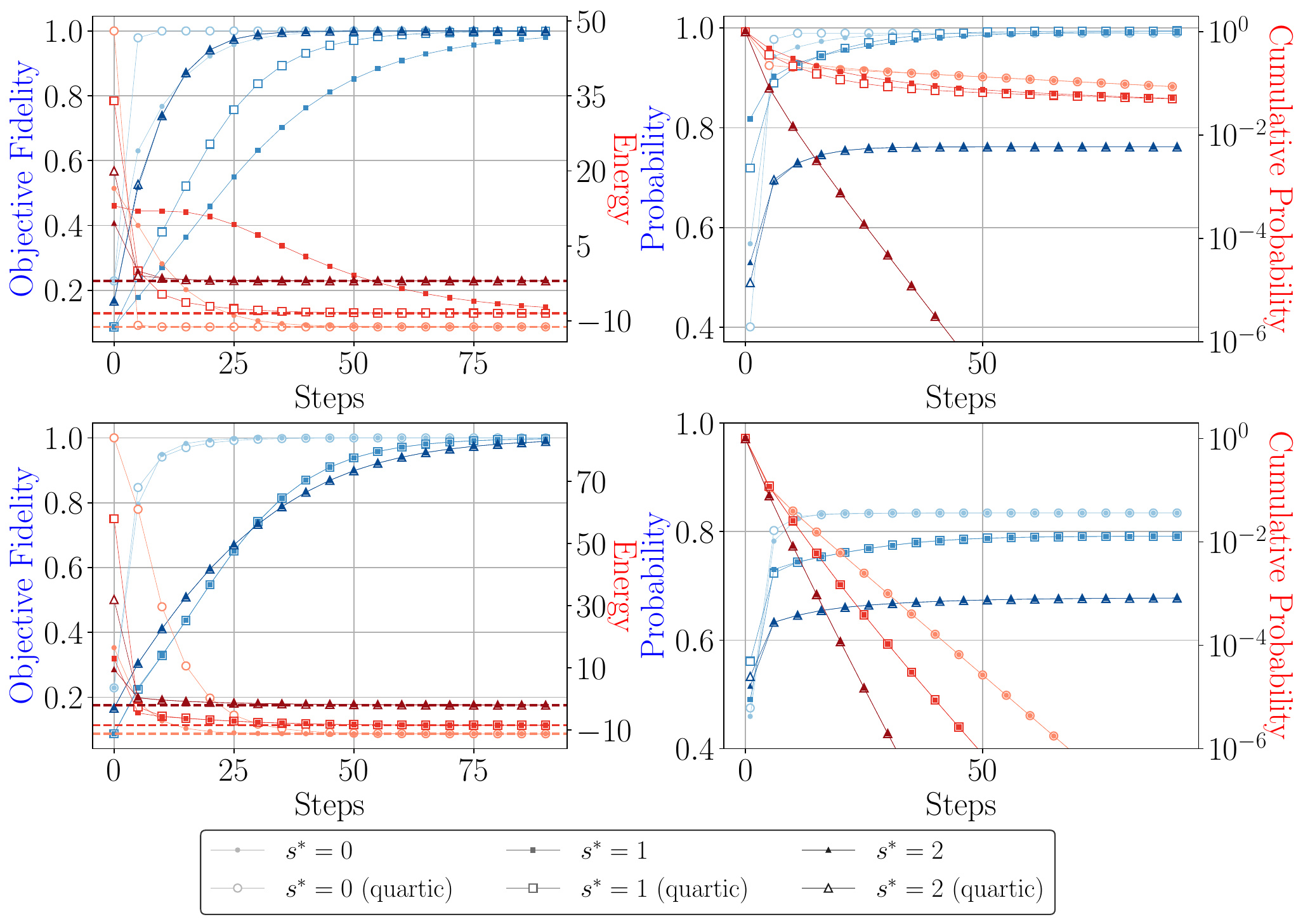}
\caption{Results comparing our proposed method with conventional penalty approach~(quartic) in PITE.
The top two figures show results of $\Delta t=0.05$, with $C'_{S}=7.5$ for our method~(filled markers) and $C_{S}=3$ for the conventional method~(open markers), while the bottom two figures show results of $\Delta t=0.015$ with $C'_{S}=7.5$ for our method, and $C_{S}=5$ for the conventional method.
``Probability" indicates the success probability of post-selection at each step, while ``Cumulative Probability" denotes the accumulated success probability up to a given step.
The dotted lines represent the exact values of $\mathcal{H}_{\mathrm{problem}}$.
}
    \label{fig:result_heisenberg_pite}
\end{figure}
\begin{figure}[tbp]
\centering
\includegraphics[width=0.8\columnwidth]{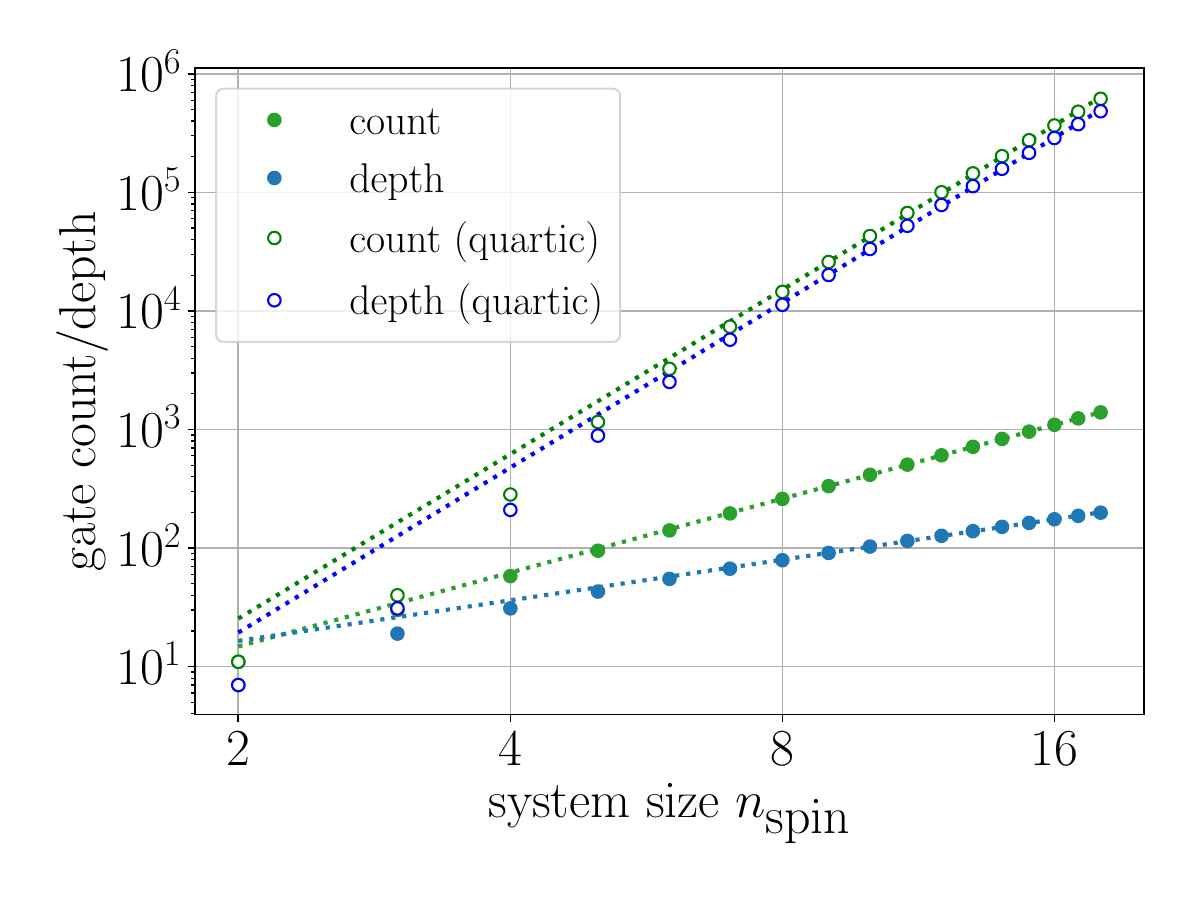}
\caption{Dependence of gate complexity of penalty terms for spin magnitude on the system size $n_{\textrm{spin}}$.
The plots were calculated by transpiling the time evolution operators, $\exp(-iC^{(')}_S\mathcal{H}_{S}^{(')}s\Delta t)=\exp(-0.15i\mathcal{H}_{S}^{(')})$, where $C^{(')}_S=7.5$, $\Delta t=0.015$, and $m_0=0.8$, into a gate set consisting of arbitrary single-qubit unitaries and CNOT gates.
The transpilation was performed using the transpile function from Qiskit~\cite{qiskit2024}.
Data points with $n_{\textrm{spin}} \geq 8$ were fitted to power laws, and the fitted curves are shown as dotted lines.
The fitting results are count: $y=3.53n_{\textrm{spin}}^{2.07}$, depth: $y=7.51n_{\textrm{spin}}^{1.14}$, count (quartic): $y=1.04n_{\textrm{spin}}^{4.61}$, and depth (quartic): $y=0.79n_{\textrm{spin}}^{4.62}$.
}
    \label{fig:gate_complexity}
\end{figure}
\begin{figure}[tbp]
    \centering
    \includegraphics[width=0.8\linewidth]{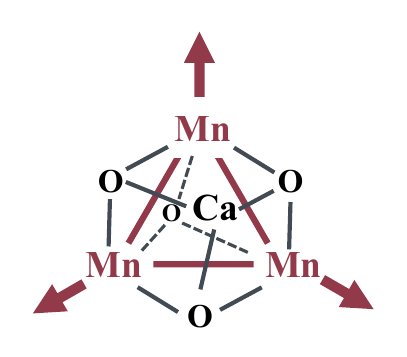}
    \caption{Schematic of a cubane-like CaMn$_3$O$_4$ cluster.}
    \label{fig:mn_figure}
\end{figure}
\begin{figure*}[tbp]
    \centering
    \includegraphics[width=\linewidth]{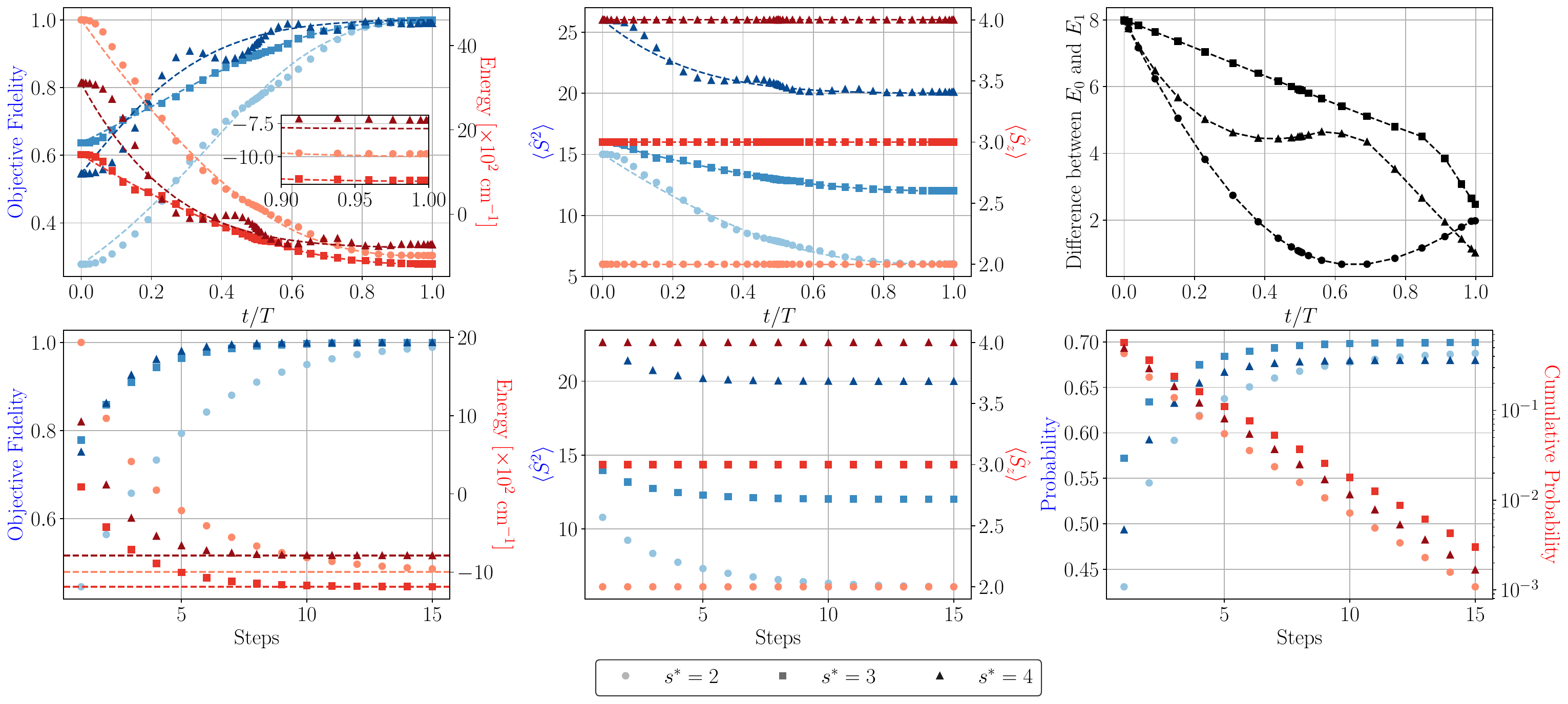}
    \caption{Results of our proposed methods for preparing spin-adapted ground states of the manganese trimer model. The top three are results of ATE, and the bottom two are results of PITE.
    The axes, labels, and notation follow the same conventions as used in \reffig{fig:result_heisenberg_adiabatic} and \reffig{fig:result_heisenberg_pite}.
    }
    \label{fig:result_mn}
\end{figure*}
\section{Numerical experiments}
In the following numerical experiments, we focus on evaluating the performance of the first stage. Accordingly, throughout this section, the desired spin-$z$ $s_z^*$ is set to $s^*$.
All numerical results in this work were obtained from classical simulations. The calculations were performed using statevector-based simulations implemented with NumPy. In Sec.~\ref{subsec:ring}, simulations were carried out using 7 qubits, and in Sec.~\ref{subsec:manganese}, 10 qubits were used.
\subsection{Spin-1/2 Heisenberg ring model}\label{subsec:ring}
In this numerical experiment, we consider the spin-1/2 Heisenberg ring model~\cite{kouzoudis1997heisenberg}.
Its Hamiltonian is given by
\begin{align}
  \mathcal{H}_{\textrm{system}}=
      \sum_{i = 0}^{n_{\textrm{spin}} - 1}
      2J_i \hat{S}_i \cdot \hat{S}_{i + 1}= \frac{1}{2}\sum_{i = 0}^{n_{\textrm{spin}} - 1} J_i \bm{\sigma}_{i}\cdot \bm{\sigma}_{i+1},
\end{align}
where the periodic boundary condition is assumed and $\bm{\sigma}_{n_{\textrm{spin}}}=\bm{\sigma}_{0}$.
In this numerical experiment, $J_i$ is fixed to $J_i=2$.

The results of $n_{\textrm{spin}}=6$ are shown in \reffig{fig:result_heisenberg_adiabatic} for ATE and \reffig{fig:result_heisenberg_pite} for PITE.
In \reffig{fig:result_heisenberg_adiabatic}, parameters are $T=1.0$, $n=2\times 10^4$, $C'_{S}=10$ and $C'_{z}=C'_{S}(2s^*+1)$.
The time schedule of $\Delta t_{i}$ is defined by a sine function as $\Delta t_{i} = \sin^2(2\pi t) \times 10^{-4}$.
The initial Hamiltonian is like $\mathcal{H}_{\textrm{initial}}=\sum_i (-1)^{i}\sigma_{i x}$ and its initial states are ones in which plus states $\ket{+}$ and minus states $\ket{-}$ are arranged alternately.
In \reffig{fig:result_heisenberg_pite}, we evaluate the convergence by comparing our proposed penalty terms with the conventional penalty term~(Eq.~(\ref{ex:conventinal_penalty})) under two different parameter settings for $C'_S$, and $\Delta t$ (see caption).
The common parameters are $m_0=0.8$ and $C'_z=C'_S(2s^*+1)$.
The initial states for the PITE simulations are $\ket{\psi}_0=\bigotimes\limits_{i=0}^{s^*+2}\ket{0}\bigotimes\limits_{i=s^{*}+3}^{n_{\mathrm{spin}}-1}\ket{1}$.
$s_z^*$ quantum numbers of the initial states are already set to the desired values $s^*$ to focus on the dependence of the spin-magnitude penalty, and they remain unchanged during the PITE calculations. 

The figures in \reffig{fig:result_heisenberg_adiabatic} demonstrate that the results of ATE successfully track the dotted lines, which represent the exact evolution based on the diagonalization of the instantaneous Hamiltonian $\mathcal{H}_t$.
In general, hyperparameters such as the time schedule $\Delta t_i$, $C'_S$, and $C'_z$ influence performance. $\Delta t$ should be chosen to be sufficiently small to ensure accuracy, whereas $C'_S$ and $C'_z$ should be taken sufficiently large to avoid significant fidelity loss.
Focusing on the deviation from the ideal energy value in the final state, we observe that it increases with $s^*$. Specifically, the energy deviation is $0.174$, $0.609$, and $1.73$ for $s^*=0$, $1$, and $2$, respectively, corresponding to fidelities of $0.996$, $0.968$, and $0.963$.
Although there is some dependence on the initial state, achieving accurate outcomes with ATE becomes increasingly challenging as the excitation level increases. 
The excitation level is determined by the energy gap between the ground state $E_0$ and the first excited state $E_1$ of the instantaneous Hamiltonian $\mathcal{H}_t$, as shown in the right panel of \reffig{fig:result_heisenberg_adiabatic}. 
Oscillatory behavior is also observed in the results. It comes from the approximation of an infinite-time schedule by a finite-time one. In fact, similar oscillations have been reported for penalty Hamiltonians for spin magnitude~\cite{sugisaki2022adiabatic}. These oscillations depend on the derivative of the time schedule at $t=0$ and $t=T$, and they occur when these derivatives are nonzero. The derivative of the time schedule in this numerical experiment becomes non-zero in the third order, which is consistent with the best-performing schedules identified by Hu~{\it et al.}~\cite{hu2016optimizing}. This suggests that our proposed method may be inherently prone to oscillatory trajectories.

The results presented in \reffig{fig:result_heisenberg_pite} demonstrate that the PITE calculations successfully converge toward the reference values denoted by the dotted lines.
In the upper two panels, the conventional penalty method (quartic) exhibits a faster convergence rate compared to our proposed method, whereas the lower panels show comparable convergence behavior between the two methods.
This indicates that the convergence properties depend on other parameters, problem settings, and the characteristics of the target states.
Conversely, when examining the gate complexity in Fig.~\ref{fig:gate_complexity}, a clear and consistent gap is observed between the two penalty approaches.
The dependencies in Fig.~\ref{fig:gate_complexity} align well with the rough complexity estimates discussed in previous sections.
Generally, it can be stated that in PITE, the cumulative success probability decreases exponentially with the increasing number of steps.
Therefore, selecting an appropriate number of steps based on the target success probability is essential for practical applications. 
The success probability at each step depends on the instantaneous quantum states; thus, preparing an initial state close to the desired state, such as by performing ATE before the PITE steps, is crucial. 
The success probabilities observed for our proposed method in the numerical experiments are comparable to those of the conventional approach and remain reasonably high in several cases, indicating that our method is practically feasible within the considered problem settings.

\subsection{Manganese Trimer}\label{subsec:manganese}
The oxygen in the atmosphere is known to be generated via the oxidation process of water
that occurs at the oxygen-evolving center (OEC) site within photosynthesis system II (PSII) \cite{bib:6813}.
The experimental results indicate the existence of a manganese cluster in the OEC.
The possible geometries of the cluster and their electronic structure, including spin states, are important for elucidating the mechanism of photosynthesis (see Refs.~\cite{bib:6763, bib:6761} and references therein).
Let us consider here a spin trimer that models the spin states of the central part, CaMn$_3$O$_4$ of the Mn cluster [See Fig.~\ref{fig:mn_figure}].
The proposed scheme in the present study is applicable to spins whose magnitudes are other than 1/2, as will be demonstrated below.

The Heisenberg Hamiltonian for the spin trimer reads
\begin{equation}
    \mathcal{H}_{\textrm{system}} = -2J_{01}\hat{S}_{0}\cdot \hat{S}_{1}-2J_{12}\hat{S}_{1}\cdot\hat{S}_{2}-2J_{20}\hat{S}_{2}\cdot\hat{S}_{0}
    ,
\end{equation}
where $\hat{S}_{i}$ represents the spin operator for a manganese ion, and $J_{ij}$ is the coefficient to represent the effective exchange interactions. In this calculation, we explore (II, II, III) set of valences in manganese. The spin magnitudes of Mn(II) and Mn(III) are $S=5/2$ and $S=2$ respectively and $J_{ij}$ values of Mn(II)-Mn(II) and Mn(II)-Mn(III) are $-1$ and $-50$ (cm$^{-1}$) respectively~\cite{bib:6761}.

In these numerical experiments, we try to obtain the $s^*=2$, $s^*=3$, and $s^*=4$ spin-adapted ground states.
For the original system Hamiltonian $\mathcal{H}_{\textrm{system}}$ the states with $s^*=3$ correspond to its ground states, those with $s^*=2$ correspond to the first excited states, and those with $s^*=4$ correspond to higher excited states above the first excited states.
The results are shown in \reffig{fig:result_mn}.
In \reffig{fig:result_mn} for ATE, the parameters are $T=1.0$, $n=2\times 10^{4}$, $C'_S=3$ and $C'_z=C'_S(2s^*+1)$.
The time schedule is the same as one of ATE for the Heisenberg Hamiltonian.
In \reffig{fig:result_mn} for PITE, parameters are 
$\Delta t=0.008$, and $m_0=0.8$.
$C'_S$ and $C'_z$ are the same as those of ATE.
For simulating the quantum system of $S=2$ and $S=5/2$, the spin state should be encoded in qubit systems.
Here, we encoded them by standard binary encoding as shown in Appendix~\ref{sec:binary}.
For the binary encoding, representing $S=2$ and $S=5/2$ states requires three qubits.
We set initial states to the computational basis like $\ket{000}\otimes\ket{000}\otimes\ket{100}$ for $s^*=2$, $\ket{000}\otimes\ket{001}\otimes\ket{100}$ for $s^*=3$ and $\ket{000}\otimes\ket{000}\otimes\ket{011}$ for $s^*=4$ where the first and second kets represent the encoded states of $S=5/2$ in Mn(II) and the third kets represent ones of $S=2$ in Mn(III).
The same initial states are used for both ATE and PITE.
The initial Hamiltonian for ATE is set according to the initial states like $\mathcal{H}_{\textrm{initial}}=\sum_{i}(-1)^{f(i)}\sigma_{iz}$ by adjusting the function of $f:\{0,1,\cdots 8\}\to\{0,1\}$.
It is worth noting that the initial states are already constrained to the desired values of $\langle \hat{S}_z\rangle$, but not of $\langle \hat{S}^2\rangle$. Throughout these calculations, only $\langle \hat{S}^2\rangle$ evolves to attain the correct spin-adapted ground states.

Both the ATE and PITE results shown in \reffig{fig:result_mn} demonstrate successful outcomes, consistent with those obtained for the Heisenberg ring model. By incorporating the proposed penalty terms, the methods can capture not only the ground state but also certain excited states. These findings indicate that our proposed methods are applicable not only to simple spin models but also to more practical and complex systems.
 
\section{Summary and Outlook}
In this work, we have proposed novel methods to reduce the computational cost of penalty terms for obtaining spin-adapted ground states of spin-rotationally symmetric Hamiltonians, from $O(n_{\textrm{spin}}^4)$ to $O(n_{\textrm{spin}}^2)$. 
To describe the practical utilities, we have shown the results for spin-1/2 Heisenberg ring models and the manganese trimer by ATE and PITE. 
For the manganese trimer, since there are few reports on the results of ATE and PITE, the novelty of this work also lies in demonstrating the feasibility of these methods for such a system.

Our main method only modifies the penalty term for spin variables, and it is widely applicable.
The preparation of excited states is generally more challenging than that of the ground state~\cite{gocho2023excited,nishi2024encoded,ma2020quantum,roggero2020preparation,nakanishi2019subspace,higgott2019variational}. However, our method enables the preparation of certain specific excited states using the same procedure as ground state preparation.
Also, it can be applied not only to various spin systems but also to first-quantized Hamiltonian~\cite{kassal2008polynomial, jones2012faster,nishiya2024first,kosugi2022imaginary,kosugi2023exhaustive}.
First-quantized formalism is considered more efficient for calculations in systems with a fixed particle number than second-quantized formalism. 
In such more complicated systems, the spin degrees of freedom can be treated independently from the spatial degrees of freedom, making them within the scope of our method.

\begin{acknowledgments}
This work was supported by the Center of Innovation for Sustainable Quantum AI, JST Grant Number JPMJPF2221.
TK acknowledges the support of JST SPRING, Grant Number JPMJSP2108.
The authors thank the Supercomputer Center, the Institute for Solid State Physics, the University of Tokyo, for the use of the facilities.
\end{acknowledgments}
\bibliographystyle{apsrev4-2}
\bibliography{ref}

@article{seki2020symmetry,
  title = {Symmetry-adapted variational quantum eigensolver},
  author = {Seki, Kazuhiro and Shirakawa, Tomonori and Yunoki, Seiji},
  journal = {Phys. Rev. A},
  volume = {101},
  issue = {5},
  pages = {052340},
  numpages = {15},
  year = {2020},
  month = {May},
  publisher = {American Physical Society},
  doi = {10.1103/PhysRevA.101.052340},
  url = {https://link.aps.org/doi/10.1103/PhysRevA.101.052340}
}

@article{nishi2023optimal,
  title = {Optimal scheduling in probabilistic imaginary-time evolution on a quantum computer},
  author = {Nishi, Hirofumi and Hamada, Koki and Nishiya, Yusuke and Kosugi, Taichi and Matsushita, Yu-ichiro},
  journal = {Phys. Rev. Res.},
  volume = {5},
  issue = {4},
  pages = {043048},
  numpages = {14},
  year = {2023},
  month = {Oct},
  publisher = {American Physical Society},
  doi = {10.1103/PhysRevResearch.5.043048},
  url = {https://link.aps.org/doi/10.1103/PhysRevResearch.5.043048}
}

@ARTICLE{meister2022resource,
       author = {{Meister}, Richard and {Benjamin}, Simon C.},
        title = "{Resource-frugal Hamiltonian eigenstate preparation via repeated quantum phase estimation measurements}",
      journal = {arXiv e-prints},
         year = 2022,
        month = dec,
          eid = {arXiv:2212.00846},
        pages = {arXiv:2212.00846},
          doi = {10.48550/arXiv.2212.00846},
archivePrefix = {arXiv},
       eprint = {2212.00846},
 primaryClass = {quant-ph},
       adsurl = {https://ui.adsabs.harvard.edu/abs/2022arXiv221200846M}
}

@article{hu2016optimizing,
  title = {Optimizing the quantum adiabatic algorithm},
  author = {Hu, Hongye and Wu, Biao},
  journal = {Phys. Rev. A},
  volume = {93},
  issue = {1},
  pages = {012345},
  numpages = {8},
  year = {2016},
  month = {Jan},
  publisher = {American Physical Society},
  doi = {10.1103/PhysRevA.93.012345},
  url = {https://link.aps.org/doi/10.1103/PhysRevA.93.012345}
}

@article{sugisaki2022adiabatic,
  author={Sugisaki, Kenji
and Toyota, Kazuo
and Sato, Kazunobu
and Shiomi, Daisuke
and Takui, Takeji},
title={Adiabatic state preparation of correlated wave functions with nonlinear scheduling functions and broken-symmetry wave functions},
journal={Communications Chemistry},
year={2022},
month={Jul},
day={25},
volume={5},
number={1},
pages={84},
issn={2399-3669},
doi={10.1038/s42004-022-00701-8},
url={https://doi.org/10.1038/s42004-022-00701-8}
}

@article{orus2019tensor,
  author={Or{\'u}s, Rom{\'a}n},
title={Tensor networks for complex quantum systems},
journal={Nature Reviews Physics},
year={2019},
month={Sep},
day={01},
volume={1},
number={9},
pages={538-550},
issn={2522-5820},
doi={10.1038/s42254-019-0086-7},
url={https://doi.org/10.1038/s42254-019-0086-7}
}

@incollection{feynman2018simulating,
  title={Simulating physics with computers},
  author={Feynman, Richard P},
  booktitle={Feynman and computation},
  pages={133--153},
  year={2018},
  publisher={cRc Press}
}

@book{nielsen2010quantum,
  title={Quantum computation and quantum information},
  author={Nielsen, Michael A and Chuang, Isaac L},
  year={2010},
  publisher={Cambridge university press}
}

@article{yoshioka2024hunting,
  author={Yoshioka, Nobuyuki
and Okubo, Tsuyoshi
and Suzuki, Yasunari
and Koizumi, Yuki
and Mizukami, Wataru},
title={Hunting for quantum-classical crossover in condensed matter problems},
journal={npj Quantum Information},
year={2024},
month={Apr},
day={29},
volume={10},
number={1},
pages={45},
issn={2056-6387},
doi={10.1038/s41534-024-00839-4},
url={https://doi.org/10.1038/s41534-024-00839-4}
}

@article{childs2018toward,
  author = {Andrew M. Childs  and Dmitri Maslov  and Yunseong Nam  and Neil J. Ross  and Yuan Su },
title = {Toward the first quantum simulation with quantum speedup},
journal = {Proceedings of the National Academy of Sciences},
volume = {115},
number = {38},
pages = {9456-9461},
year = {2018},
doi = {10.1073/pnas.1801723115},
URL = {https://www.pnas.org/doi/abs/10.1073/pnas.1801723115},
}

@article{georgescu2014quantum,
  title = {Quantum simulation},
  author = {Georgescu, I. M. and Ashhab, S. and Nori, Franco},
  journal = {Rev. Mod. Phys.},
  volume = {86},
  issue = {1},
  pages = {153--185},
  numpages = {33},
  year = {2014},
  month = {Mar},
  publisher = {American Physical Society},
  doi = {10.1103/RevModPhys.86.153},
  url = {https://link.aps.org/doi/10.1103/RevModPhys.86.153}
}

@article{childs2021theory,
   title = {Theory of {T}rotter {E}rror with {C}ommutator {S}caling},
  author = {Childs, Andrew M. and Su, Yuan and Tran, Minh C. and Wiebe, Nathan and Zhu, Shuchen},
  journal = {Phys. Rev. X},
  volume = {11},
  issue = {1},
  pages = {011020},
  numpages = {49},
  year = {2021},
  month = {Feb},
  publisher = {American Physical Society},
  doi = {10.1103/PhysRevX.11.011020},
  url = {https://link.aps.org/doi/10.1103/PhysRevX.11.011020}
}

@inproceedings{sandvik2010computational,
  title={Computational studies of quantum spin systems},
  author={Sandvik, Anders W},
  booktitle={AIP Conference Proceedings},
  volume={1297},
  number={1},
  pages={135--338},
  year={2010},
  organization={American Institute of Physics},
  url={https://doi.org/10.1063/1.3518900}
}

@article{lyu2023symmetry,
  title={Symmetry enhanced variational quantum spin eigensolver},
  author={Lyu, Chufan and Xu, Xusheng and Yung, Man-Hong and Bayat, Abolfazl},
  journal={Quantum},
  volume={7},
  pages={899},
  year={2023},
  publisher={Verein zur F{\"o}rderung des Open Access Publizierens in den Quantenwissenschaften},
  url={https://doi.org/10.22331/q-2023-01-19-899}
}

@ARTICLE{farhi2000quantum,
       author = {{Farhi}, Edward and {Goldstone}, Jeffrey and {Gutmann}, Sam and {Sipser}, Michael},
        title = "{Quantum Computation by Adiabatic Evolution}",
      journal = {arXiv e-prints},
         year = 2000,
        month = jan,
          eid = {quant-ph/0001106},
        pages = {quant-ph/0001106},
          doi = {10.48550/arXiv.quant-ph/0001106},
archivePrefix = {arXiv},
       eprint = {quant-ph/0001106},
 primaryClass = {quant-ph},
       adsurl = {https://ui.adsabs.harvard.edu/abs/2000quant.ph..1106F}
}

@article{hejazi2024adiabatic,
  title = {Adiabatic quantum imaginary time evolution},
  author = {Hejazi, Kasra and Motta, Mario and Chan, Garnet Kin-Lic},
  journal = {Phys. Rev. Res.},
  volume = {6},
  issue = {3},
  pages = {033084},
  numpages = {7},
  year = {2024},
  month = {Jul},
  publisher = {American Physical Society},
  doi = {10.1103/PhysRevResearch.6.033084},
  url = {https://link.aps.org/doi/10.1103/PhysRevResearch.6.033084}
}

@article{sun2021quantum,
  title = {Quantum {C}omputation of {F}inite-{T}emperature {S}tatic and {D}ynamical {P}roperties of {S}pin {S}ystems {U}sing {Q}uantum {I}maginary {T}ime {E}volution},
  author = {Sun, Shi-Ning and Motta, Mario and Tazhigulov, Ruslan N. and Tan, Adrian T.K. and Chan, Garnet Kin-Lic and Minnich, Austin J.},
  journal = {PRX Quantum},
  volume = {2},
  issue = {1},
  pages = {010317},
  numpages = {14},
  year = {2021},
  month = {Feb},
  publisher = {American Physical Society},
  doi = {10.1103/PRXQuantum.2.010317},
  url = {https://link.aps.org/doi/10.1103/PRXQuantum.2.010317}
}

@article{kosugi2022imaginary,
  title = {Imaginary-time evolution using forward and backward real-time evolution with a single ancilla: {F}irst-quantized eigensolver algorithm for quantum chemistry},
  author = {Kosugi, Taichi and Nishiya, Yusuke and Nishi, Hirofumi and Matsushita, Yu-ichiro},
  journal = {Phys. Rev. Res.},
  volume = {4},
  issue = {3},
  pages = {033121},
  numpages = {13},
  year = {2022},
  month = {Aug},
  publisher = {American Physical Society},
  doi = {10.1103/PhysRevResearch.4.033121},
  url = {https://link.aps.org/doi/10.1103/PhysRevResearch.4.033121}
}

@article{kadowaki1998quantum,
  title = {Quantum annealing in the transverse {I}sing model},
  author = {Kadowaki, Tadashi and Nishimori, Hidetoshi},
  journal = {Phys. Rev. E},
  volume = {58},
  issue = {5},
  pages = {5355--5363},
  numpages = {0},
  year = {1998},
  month = {Nov},
  publisher = {American Physical Society},
  doi = {10.1103/PhysRevE.58.5355},
  url = {https://link.aps.org/doi/10.1103/PhysRevE.58.5355}
}

@article{bib:6761,
title = {Electronic and spin structures of manganese clusters in the photosynthesis {II} system},
journal = {Polyhedron},
volume = {24},
number = {16},
pages = {2767-2777},
year = {2005},
note = {Proceedings of the 9th International Conference on Molecule-based Magnets (ICMM 2004)},
issn = {0277-5387},
doi = {https://doi.org/10.1016/j.poly.2005.08.049},
url = {https://www.sciencedirect.com/science/article/pii/S0277538705005620},
author = {Hiroshi Isobe and Mitsuo Shoji and Kennichi Koizumi and Yasutaka Kitagawa and Shusuke Yamanaka and Seiki Kuramitsu and Kizashi Yamaguchi},
}

@Article{bib:6763,
author={Hendrickson, David N.
and Christou, George
and Schmitt, Edward A.
and Libby, Eduardo
and Bashkin, John S.
and Wang, Sheyi
and Tsai, Hui Lien
and Vincent, John B.
and Boyd, Peter D. W.},
title={Photosynthetic water oxidation center: spin frustration in distorted cubane {MnIVMnIII3} model complexes},
journal={Journal of the American Chemical Society},
year={1992},
month={Mar},
day={01},
publisher={American Chemical Society},
volume={114},
number={7},
pages={2455-2471},
issn={0002-7863},
doi={10.1021/ja00033a022},
url={https://doi.org/10.1021/ja00033a022}
}

@article{bib:6813,
author = {Kristina N. Ferreira  and Tina M. Iverson  and Karim Maghlaoui  and James Barber  and So Iwata },
title = {Architecture of the {P}hotosynthetic {O}xygen-{E}volving Center},
journal = {Science},
volume = {303},
number = {5665},
pages = {1831-1838},
year = {2004},
doi = {10.1126/science.1093087},
url = {https://www.science.org/doi/abs/10.1126/science.1093087},
}

@article{sawaya2020resource,
  author={Sawaya, Nicolas P. D.
and Menke, Tim
and Kyaw, Thi Ha
and Johri, Sonika
and Aspuru-Guzik, Al{\'a}n
and Guerreschi, Gian Giacomo},
title={Resource-efficient digital quantum simulation of d-level systems for photonic, vibrational, and spin-s {H}amiltonians},
journal={npj Quantum Information},
year={2020},
month={Jun},
day={02},
volume={6},
number={1},
pages={49},
issn={2056-6387},
doi={10.1038/s41534-020-0278-0},
url={https://doi.org/10.1038/s41534-020-0278-0}
}

@article{kassal2008polynomial,
  author = {Ivan Kassal  and Stephen P. Jordan  and Peter J. Love  and Masoud Mohseni  and Alán Aspuru-Guzik },
title = {Polynomial-time quantum algorithm for the simulation of chemical dynamics},
journal = {Proceedings of the National Academy of Sciences},
volume = {105},
number = {48},
pages = {18681-18686},
year = {2008},
doi = {10.1073/pnas.0808245105},
URL = {https://www.pnas.org/doi/abs/10.1073/pnas.0808245105},
}

@article{jones2012faster,
  doi = {10.1088/1367-2630/14/11/115023},
url = {https://dx.doi.org/10.1088/1367-2630/14/11/115023},
year = {2012},
month = {nov},
publisher = {IOP Publishing},
volume = {14},
number = {11},
pages = {115023},
author = {N Cody Jones and James D Whitfield and Peter L McMahon and Man-Hong Yung and Rodney Van Meter and Alán Aspuru-Guzik and Yoshihisa Yamamoto},
title = {Faster quantum chemistry simulation on fault-tolerant quantum computers},
journal = {New Journal of Physics},
}

@article{nishiya2024first,
  title = {First-quantized adiabatic time evolution for the ground state of a many-electron system and the optimal nuclear configuration},
  author = {Nishiya, Yusuke and Nishi, Hirofumi and Couzini\'e, Yannick and Kosugi, Taichi and Matsushita, Yu-ichiro},
  journal = {Phys. Rev. A},
  volume = {109},
  issue = {2},
  pages = {022423},
  numpages = {10},
  year = {2024},
  month = {Feb},
  publisher = {American Physical Society},
  doi = {10.1103/PhysRevA.109.022423},
  url = {https://link.aps.org/doi/10.1103/PhysRevA.109.022423}
}

@article{kosugi2023exhaustive,
  author={Kosugi, Taichi
and Nishi, Hirofumi
and Matsushita, Yu-ichiro},
title={Exhaustive search for optimal molecular geometries using imaginary-time evolution on a quantum computer},
journal={npj Quantum Information},
year={2023},
month={Nov},
day={03},
volume={9},
number={1},
pages={112},
issn={2056-6387},
doi={10.1038/s41534-023-00778-6},
url={https://doi.org/10.1038/s41534-023-00778-6}
}

@article{keller2016spin,
  author={Keller, Sebastian
and Reiher, Markus},
title={Spin-adapted matrix product states and operators},
journal={The Journal of Chemical Physics},
year={2016},
month={Apr},
day={01},
volume={144},
number={13},
pages={134101},
issn={0021-9606},
doi={10.1063/1.4944921},
url={https://doi.org/10.1063/1.4944921}
}

@article{sharma2012spin,
  author={Sharma, Sandeep
and Chan, Garnet Kin-Lic},
title={Spin-adapted density matrix renormalization group algorithms for quantum chemistry},
journal={The Journal of Chemical Physics},
year={2012},
month={Mar},
day={30},
volume={136},
number={12},
pages={124121},
issn={0021-9606},
doi={10.1063/1.3695642},
url={https://doi.org/10.1063/1.3695642}
}

@article{gocho2023excited,
  author={Gocho, Shigeki
and Nakamura, Hajime
and Kanno, Shu
and Gao, Qi
and Kobayashi, Takao
and Inagaki, Taichi
and Hatanaka, Miho},
title={Excited state calculations using variational quantum eigensolver with spin-restricted ans{\"a}tze and automatically-adjusted constraints},
journal={npj Computational Materials},
year={2023},
month={Jan},
day={21},
volume={9},
number={1},
pages={13},
issn={2057-3960},
doi={10.1038/s41524-023-00965-1},
url={https://doi.org/10.1038/s41524-023-00965-1}
}

@ARTICLE{nishi2024encoded,
       author = {{Nishi}, Hirofumi and {Takei}, Yuki and {Kosugi}, Taichi and {Mieda}, Shunsuke and {Natsume}, Yutaka and {Aoyagi}, Takeshi and {Matsushita}, Yu-ichiro},
        title = "{Encoded probabilistic imaginary-time evolution on a trapped-ion quantum computer for ground and excited states of spin qubits}",
      journal = {arXiv e-prints},
         year = 2024,
        month = jul,
          eid = {arXiv:2407.10555},
        pages = {arXiv:2407.10555},
          doi = {10.48550/arXiv.2407.10555},
archivePrefix = {arXiv},
       eprint = {2407.10555},
 primaryClass = {quant-ph},
       adsurl = {https://ui.adsabs.harvard.edu/abs/2024arXiv240710555N}
}

@article{ma2020quantum,
  author={Ma, He
and Govoni, Marco
and Galli, Giulia},
title={Quantum simulations of materials on near-term quantum computers},
journal={npj Computational Materials},
year={2020},
month={Jul},
day={02},
volume={6},
number={1},
pages={85},
issn={2057-3960},
doi={10.1038/s41524-020-00353-z},
url={https://doi.org/10.1038/s41524-020-00353-z}
}

@article{roggero2020preparation,
  title = {Preparation of excited states for nuclear dynamics on a quantum computer},
  author = {Roggero, Alessandro and Gu, Chenyi and Baroni, Alessandro and Papenbrock, Thomas},
  journal = {Phys. Rev. C},
  volume = {102},
  issue = {6},
  pages = {064624},
  numpages = {23},
  year = {2020},
  month = {Dec},
  publisher = {American Physical Society},
  doi = {10.1103/PhysRevC.102.064624},
  url = {https://link.aps.org/doi/10.1103/PhysRevC.102.064624}
}

@article{kuroiwa2021penalty,
  title = {Penalty methods for a variational quantum eigensolver},
  author = {Kuroiwa, Kohdai and Nakagawa, Yuya O.},
  journal = {Phys. Rev. Res.},
  volume = {3},
  issue = {1},
  pages = {013197},
  numpages = {11},
  year = {2021},
  month = {Feb},
  publisher = {American Physical Society},
  doi = {10.1103/PhysRevResearch.3.013197},
  url = {https://link.aps.org/doi/10.1103/PhysRevResearch.3.013197}
}

@article{lowdin1969exchange,
  author = {Löwdin, Per-Olov and Goscinski, Osvaldo},
title = {The exchange phenomenon, the symmetric group, and the spin degeneracy problem},
journal = {International Journal of Quantum Chemistry},
volume = {4},
number = {S3B},
pages = {533-591},
doi = {https://doi.org/10.1002/qua.560040719},
url = {https://onlinelibrary.wiley.com/doi/abs/10.1002/qua.560040719},
year = {1969}
}

@book{wigner2012group,
  title={Group Theory: And its Application to the Quantum Mechanics of Atomic Spectra},
  author={Wigner, E.},
  isbn={9780323152785},
  series={Pure and applied physics},
  url={https://books.google.co.jp/books?id=ENZzI49uZMcC},
  year={2012},
  publisher={Elsevier Science}
}

@ARTICLE{zi2024shallow,
       author = {{Zi}, Wei and {Nie}, Junhong and {Sun}, Xiaoming},
        title = "{Shallow Quantum Circuit Implementation of Symmetric Functions with Limited Ancillary Qubits}",
      journal = {arXiv e-prints},
         year = 2024,
        month = apr,
          eid = {arXiv:2404.06052},
        pages = {arXiv:2404.06052},
          doi = {10.48550/arXiv.2404.06052},
archivePrefix = {arXiv},
       eprint = {2404.06052},
 primaryClass = {quant-ph},
       adsurl = {https://ui.adsabs.harvard.edu/abs/2024arXiv240406052Z}
}

@article{carbone2022quantum,
  AUTHOR = {Carbone, Alessandro and Galli, Davide Emilio and Motta, Mario and Jones, Barbara},
TITLE = {Quantum {C}ircuits for the {P}reparation of {S}pin {E}igenfunctions on {Q}uantum {C}omputers},
JOURNAL = {Symmetry},
VOLUME = {14},
YEAR = {2022},
NUMBER = {3},
ARTICLE-NUMBER = {624},
URL = {https://www.mdpi.com/2073-8994/14/3/624},
ISSN = {2073-8994},
DOI = {10.3390/sym14030624}

}

@book{pauncz2012spin,
  title={Spin eigenfunctions: construction and use},
  author={Pauncz, Ruben},
  year={2012},
  publisher={Springer Science \& Business Media}
}

@article{xie2024probabilistic,
  title = {Probabilistic imaginary-time evolution algorithm based on nonunitary quantum circuits},
  author = {Xie, Hao-Nan and Wei, Shi-Jie and Yang, Fan and Wang, Zheng-An and Chen, Chi-Tong and Fan, Heng and Long, Gui-Lu},
  journal = {Phys. Rev. A},
  volume = {109},
  issue = {5},
  pages = {052414},
  numpages = {15},
  year = {2024},
  month = {May},
  publisher = {American Physical Society},
  doi = {10.1103/PhysRevA.109.052414},
  url = {https://link.aps.org/doi/10.1103/PhysRevA.109.052414}
}

@article{liu2021probabilistic,
  author={Liu, Tong
and Liu, Jin-Guo
and Fan, Heng},
title={Probabilistic nonunitary gate in imaginary time evolution},
journal={Quantum Information Processing},
year={2021},
month={Jun},
day={08},
volume={20},
number={6},
pages={204},
issn={1573-1332},
doi={10.1007/s11128-021-03145-6},
url={https://doi.org/10.1007/s11128-021-03145-6}
}

@article{nishi2024quadratic,
  title = {Quadratic acceleration of multistep probabilistic algorithms for state preparation},
  author = {Nishi, Hirofumi and Kosugi, Taichi and Nishiya, Yusuke and Matsushita, Yu-ichiro},
  journal = {Phys. Rev. Res.},
  volume = {6},
  issue = {2},
  pages = {L022041},
  numpages = {6},
  year = {2024},
  month = {May},
  publisher = {American Physical Society},
  doi = {10.1103/PhysRevResearch.6.L022041},
  url = {https://link.aps.org/doi/10.1103/PhysRevResearch.6.L022041}
}

@article{lloyd1996universal,
  author = {Seth Lloyd },
title = {Universal {Q}uantum {S}imulators},
journal = {Science},
volume = {273},
number = {5278},
pages = {1073-1078},
year = {1996},
doi = {10.1126/science.273.5278.1073},
URL = {https://www.science.org/doi/abs/10.1126/science.273.5278.1073},
}

@article{ivan2008polynomial,
author = {Ivan Kassal  and Stephen P. Jordan  and Peter J. Love  and Masoud Mohseni  and Alán Aspuru-Guzik },
title = {Polynomial-time quantum algorithm for the simulation of chemical dynamics},
journal = {Proceedings of the National Academy of Sciences},
volume = {105},
number = {48},
pages = {18681-18686},
year = {2008},
doi = {10.1073/pnas.0808245105},
url = {https://www.pnas.org/doi/abs/10.1073/pnas.0808245105}}

@book{kitaev2002classical,
  title={Classical and quantum computation},
  author={Kitaev, Alexei Yu and Shen, Alexander and Vyalyi, Mikhail N},
  number={47},
  year={2002},
  publisher={American Mathematical Soc.}
}

@article{kouzoudis1997heisenberg,
title = {Heisenberg $s = \frac{1}{2}$ ring consisting of a prime number of atoms},
journal = {Journal of Magnetism and Magnetic Materials},
volume = {173},
number = {3},
pages = {259-265},
year = {1997},
issn = {0304-8853},
doi = {https://doi.org/10.1016/S0304-8853(97)00234-5},
url = {https://www.sciencedirect.com/science/article/pii/S0304885397002345},
author = {D. Kouzoudis},
}

@article{low2019hamiltonian,
  doi = {10.22331/q-2019-07-12-163},
  url = {https://doi.org/10.22331/q-2019-07-12-163},
  title = {Hamiltonian {S}imulation by {Q}ubitization},
  author = {Low, Guang Hao and Chuang, Isaac L.},
  journal = {{Quantum}},
  issn = {2521-327X},
  publisher = {{Verein zur F{\"{o}}rderung des Open Access Publizierens in den Quantenwissenschaften}},
  volume = {3},
  pages = {163},
  month = jul,
  year = {2019}
}

@article{bauer2020quantum,
  author={Bauer, Bela
and Bravyi, Sergey
and Motta, Mario
and Chan, Garnet Kin-Lic},
title={Quantum {A}lgorithms for {Q}uantum {C}hemistry and {Q}uantum {M}aterials {S}cience},
journal={Chemical Reviews},
year={2020},
month={Nov},
day={25},
publisher={American Chemical Society},
volume={120},
number={22},
pages={12685-12717},
issn={0009-2665},
doi={10.1021/acs.chemrev.9b00829},
url={https://doi.org/10.1021/acs.chemrev.9b00829}
}

@article{cerezo2021variational,
  author={Cerezo, M.
and Arrasmith, Andrew
and Babbush, Ryan
and Benjamin, Simon C.
and Endo, Suguru
and Fujii, Keisuke
and McClean, Jarrod R.
and Mitarai, Kosuke
and Yuan, Xiao
and Cincio, Lukasz
and Coles, Patrick J.},
title={Variational quantum algorithms},
journal={Nature Reviews Physics},
year={2021},
month={Sep},
day={01},
volume={3},
number={9},
pages={625-644},
issn={2522-5820},
doi={10.1038/s42254-021-00348-9},
url={https://doi.org/10.1038/s42254-021-00348-9}
}

@article{cao2019quantum,
  author={Cao, Yudong
and Romero, Jonathan
and Olson, Jonathan P.
and Degroote, Matthias
and Johnson, Peter D.
and Kieferov{\'a}, M{\'a}ria
and Kivlichan, Ian D.
and Menke, Tim
and Peropadre, Borja
and Sawaya, Nicolas P. D.
and Sim, Sukin
and Veis, Libor
and Aspuru-Guzik, Al{\'a}n},
title={Quantum {C}hemistry in the {A}ge of {Q}uantum {C}omputing},
journal={Chemical Reviews},
year={2019},
month={Oct},
day={09},
publisher={American Chemical Society},
volume={119},
number={19},
pages={10856-10915},
issn={0009-2665},
doi={10.1021/acs.chemrev.8b00803},
url={https://doi.org/10.1021/acs.chemrev.8b00803}
}

@article{jahnke2000error,
  author={Jahnke, Tobias
and Lubich, Christian},
title={Error {B}ounds for {E}xponential {O}perator {S}plittings},
journal={BIT Numerical Mathematics},
year={2000},
month={Dec},
day={01},
volume={40},
number={4},
pages={735-744},
issn={1572-9125},
doi={10.1023/A:1022396519656},
url={https://doi.org/10.1023/A:1022396519656}
}

@article{siwach2021filtering,
  title = {Filtering states with total spin on a quantum computer},
  author = {Siwach, Pooja and Lacroix, Denis},
  journal = {Phys. Rev. A},
  volume = {104},
  issue = {6},
  pages = {062435},
  numpages = {9},
  year = {2021},
  month = {Dec},
  publisher = {American Physical Society},
  doi = {10.1103/PhysRevA.104.062435},
  url = {https://link.aps.org/doi/10.1103/PhysRevA.104.062435}
}

@article{stetcu2023projection,
  title = {Projection algorithm for state preparation on quantum computers},
  author = {Stetcu, I. and Baroni, A. and Carlson, J.},
  journal = {Phys. Rev. C},
  volume = {108},
  issue = {3},
  pages = {L031306},
  numpages = {6},
  year = {2023},
  month = {Sep},
  publisher = {American Physical Society},
  doi = {10.1103/PhysRevC.108.L031306},
  url = {https://link.aps.org/doi/10.1103/PhysRevC.108.L031306}
}

@article{lacroix2020symmetry,
  title = {Symmetry-{A}ssisted {P}reparation of {E}ntangled {M}any-{B}ody {S}tates on a {Q}uantum {C}omputer},
  author = {Lacroix, Denis},
  journal = {Phys. Rev. Lett.},
  volume = {125},
  issue = {23},
  pages = {230502},
  numpages = {6},
  year = {2020},
  month = {Dec},
  publisher = {American Physical Society},
  doi = {10.1103/PhysRevLett.125.230502},
  url = {https://link.aps.org/doi/10.1103/PhysRevLett.125.230502}
}

@article{nakanishi2019subspace,
  title = {Subspace-search variational quantum eigensolver for excited states},
  author = {Nakanishi, Ken M. and Mitarai, Kosuke and Fujii, Keisuke},
  journal = {Phys. Rev. Res.},
  volume = {1},
  issue = {3},
  pages = {033062},
  numpages = {7},
  year = {2019},
  month = {Oct},
  publisher = {American Physical Society},
  doi = {10.1103/PhysRevResearch.1.033062},
  url = {https://link.aps.org/doi/10.1103/PhysRevResearch.1.033062}
}

@article{higgott2019variational,
   doi = {10.22331/q-2019-07-01-156},
  url = {https://doi.org/10.22331/q-2019-07-01-156},
  title = {Variational {Q}uantum {C}omputation of {E}xcited {S}tates},
  author = {Higgott, Oscar and Wang, Daochen and Brierley, Stephen},
  journal = {{Quantum}},
  issn = {2521-327X},
  publisher = {{Verein zur F{\"{o}}rderung des Open Access Publizierens in den Quantenwissenschaften}},
  volume = {3},
  pages = {156},
  month = jul,
  year = {2019}
}

@article{mitarai2018quantum,
  title = {Quantum circuit learning},
  author = {Mitarai, K. and Negoro, M. and Kitagawa, M. and Fujii, K.},
  journal = {Phys. Rev. A},
  volume = {98},
  issue = {3},
  pages = {032309},
  numpages = {6},
  year = {2018},
  month = {Sep},
  publisher = {American Physical Society},
  doi = {10.1103/PhysRevA.98.032309},
  url = {https://link.aps.org/doi/10.1103/PhysRevA.98.032309}
}

@article{motta2020determining,
  author={Motta, Mario
and Sun, Chong
and Tan, Adrian T. K.
and O'Rourke, Matthew J.
and Ye, Erika
and Minnich, Austin J.
and Brand{\~a}o, Fernando G. S. L.
and Chan, Garnet Kin-Lic},
title={Determining eigenstates and thermal states on a quantum computer using quantum imaginary time evolution},
journal={Nature Physics},
year={2020},
month={Feb},
day={01},
volume={16},
number={2},
pages={205-210},
issn={1745-2481},
doi={10.1038/s41567-019-0704-4},
url={https://doi.org/10.1038/s41567-019-0704-4}
}

@article{trotter1959product,
  ISSN = {00029939, 10886826},
 URL = {http://www.jstor.org/stable/2033649},
 author = {H. F. Trotter},
 journal = {Proceedings of the American Mathematical Society},
 number = {4},
 pages = {545--551},
 publisher = {American Mathematical Society},
 title = {On the {P}roduct of {S}emi-{G}roups of {O}perators},
 urldate = {2024-12-25},
 volume = {10},
 year = {1959}
}

@article{suzuki1976relationship,
  author={Suzuki, Masuo},
title={Relationship between d-{D}imensional {Q}uantal {S}pin {S}ystems and (d+1)-{D}imensional {I}sing {S}ystems: {E}quivalence, {C}ritical {E}xponents and {S}ystematic {A}pproximants of the {P}artition {F}unction and {S}pin {C}orrelations},
journal={Progress of Theoretical Physics},
year={1976},
month={Nov},
day={01},
volume={56},
number={5},
pages={1454-1469},
issn={0033-068X},
doi={10.1143/PTP.56.1454},
url={https://doi.org/10.1143/PTP.56.1454}
}

@article{wu2024variational,
  author={Wu, Dian
and Rossi, Riccardo
and Vicentini, Filippo
and Astrakhantsev, Nikita
and Becca, Federico
and Cao, Xiaodong
and Carrasquilla, Juan
and Ferrari, Francesco
and Georges, Antoine
and Hibat-Allah, Mohamed
and Imada, Masatoshi
and L{\"a}uchli, Andreas M.
and Mazzola, Guglielmo
and Mezzacapo, Antonio
and Millis, Andrew
and Robledo Moreno, Javier
and Neupert, Titus
and Nomura, Yusuke
and Nys, Jannes
and Parcollet, Olivier
and Pohle, Rico
and Romero, Imelda
and Schmid, Michael
and Silvester, J. Maxwell
and Sorella, Sandro
and Tocchio, Luca F.
and Wang, Lei
and White, Steven R.
and Wietek, Alexander
and Yang, Qi
and Yang, Yiqi
and Zhang, Shiwei
and Carleo, Giuseppe},
title={Variational benchmarks for quantum many-body problems},
journal={Science},
year={2024},
month={Oct},
day={18},
publisher={American Association for the Advancement of Science},
volume={386},
number={6719},
pages={296-301},
doi={10.1126/science.adg9774},
url={https://doi.org/10.1126/science.adg9774}
}

@article{lacroix2023symmetry,
  author={Lacroix, Denis
and Ruiz Guzman, Edgar Andres
and Siwach, Pooja},
title={Symmetry breaking/symmetry preserving circuits and symmetry restoration on quantum computers},
journal={The European Physical Journal A},
year={2023},
month={Jan},
day={10},
volume={59},
number={1},
pages={3},
issn={1434-601X},
doi={10.1140/epja/s10050-022-00911-7},
url={https://doi.org/10.1140/epja/s10050-022-00911-7}
}

@article{jones2019variational,
  title = {Variational quantum algorithms for discovering {H}amiltonian spectra},
  author = {Jones, Tyson and Endo, Suguru and McArdle, Sam and Yuan, Xiao and Benjamin, Simon C.},
  journal = {Phys. Rev. A},
  volume = {99},
  issue = {6},
  pages = {062304},
  numpages = {9},
  year = {2019},
  month = {Jun},
  publisher = {American Physical Society},
  doi = {10.1103/PhysRevA.99.062304},
  url = {https://link.aps.org/doi/10.1103/PhysRevA.99.062304}
}

@article{mcardle2019variational,
  author={McArdle, Sam
and Jones, Tyson
and Endo, Suguru
and Li, Ying
and Benjamin, Simon C.
and Yuan, Xiao},
title={Variational ansatz-based quantum simulation of imaginary time evolution},
journal={npj Quantum Information},
year={2019},
month={Sep},
day={06},
volume={5},
number={1},
pages={75},
issn={2056-6387},
doi={10.1038/s41534-019-0187-2},
url={https://doi.org/10.1038/s41534-019-0187-2}
}

@article{yuan2019theory,
 doi = {10.22331/q-2019-10-07-191},
  url = {https://doi.org/10.22331/q-2019-10-07-191},
  title = {Theory of variational quantum simulation},
  author = {Yuan, Xiao and Endo, Suguru and Zhao, Qi and Li, Ying and Benjamin, Simon C.},
  journal = {{Quantum}},
  issn = {2521-327X},
  publisher = {{Verein zur F{\"{o}}rderung des Open Access Publizierens in den Quantenwissenschaften}},
  volume = {3},
  pages = {191},
  month = oct,
  year = {2019}
}

@article{yeter2020practical,
  author={Yeter-Aydeniz, K{\"u}bra
and Pooser, Raphael C.
and Siopsis, George},
title={Practical quantum computation of chemical and nuclear energy levels using quantum imaginary time evolution and Lanczos algorithms},
journal={npj Quantum Information},
year={2020},
month={Jul},
day={17},
volume={6},
number={1},
pages={63},
issn={2056-6387},
doi={10.1038/s41534-020-00290-1},
url={https://doi.org/10.1038/s41534-020-00290-1}
}

@article{lin2021real,
  title = {Real- and {I}maginary-{T}ime {E}volution with {C}ompressed {Q}uantum {C}ircuits},
  author = {Lin, Sheng-Hsuan and Dilip, Rohit and Green, Andrew G. and Smith, Adam and Pollmann, Frank},
  journal = {PRX Quantum},
  volume = {2},
  issue = {1},
  pages = {010342},
  numpages = {15},
  year = {2021},
  month = {Mar},
  publisher = {American Physical Society},
  doi = {10.1103/PRXQuantum.2.010342},
  url = {https://link.aps.org/doi/10.1103/PRXQuantum.2.010342}
}

@ARTICLE{gandon2024quantumcomputingspinadaptedrepresentations,
       author = {{Gandon}, Anthony and {Baiardi}, Alberto and {Rossmannek}, Max and {Dobrautz}, Werner and {Tavernelli}, Ivano},
        title = "{Quantum computing in spin-adapted representations for efficient simulations of spin systems}",
      journal = {arXiv e-prints},
         year = 2024,
        month = dec,
          eid = {arXiv:2412.14797},
        pages = {arXiv:2412.14797},
          doi = {10.48550/arXiv.2412.14797},
archivePrefix = {arXiv},
       eprint = {2412.14797},
 primaryClass = {quant-ph},
       adsurl = {https://ui.adsabs.harvard.edu/abs/2024arXiv241214797G}
}

@Article{Planelles1994,
author={Planelles, Josep
and Viciano, Pascual},
title={Reduced Hamiltonians. I: Spin-adapted and spin-nonadapted reduced Hamiltonians},
journal={Journal of Mathematical Chemistry},
year={1994},
month={Dec},
day={01},
volume={16},
number={1},
pages={137-148},
issn={1572-8897},
doi={10.1007/BF01169202},
url={https://doi.org/10.1007/BF01169202}
}

@book{stanley2015catalan,
  title={Catalan numbers},
  author={Stanley, Richard P},
  year={2015},
  publisher={Cambridge University Press}
}

@article{moudgalya2028entanglement,
  title = {Entanglement of exact excited states of Affleck-Kennedy-Lieb-Tasaki models: Exact results, many-body scars, and violation of the strong eigenstate thermalization hypothesis},
  author = {Moudgalya, Sanjay and Regnault, Nicolas and Bernevig, B. Andrei},
  journal = {Phys. Rev. B},
  volume = {98},
  issue = {23},
  pages = {235156},
  numpages = {43},
  year = {2018},
  month = {Dec},
  publisher = {American Physical Society},
  doi = {10.1103/PhysRevB.98.235156},
  url = {https://link.aps.org/doi/10.1103/PhysRevB.98.235156}
}

@book{essler2005one,
  title={The one-dimensional Hubbard model},
  author={Essler, Fabian HL and Frahm, Holger and G{\"o}hmann, Frank and Kl{\"u}mper, Andreas and Korepin, Vladimir E},
  year={2005},
  publisher={Cambridge University Press}
}

@article{chilton2022molecular,
   author = "Chilton, Nicholas F.",
   title = "Molecular Magnetism", 
   journal= "Annual Review of Materials Research",
   year = "2022",
   volume = "52",
   number = "Volume 52, 2022",
   pages = "79-101",
   doi = "https://doi.org/10.1146/annurev-matsci-081420-042553",
   url = "https://www.annualreviews.org/content/journals/10.1146/annurev-matsci-081420-042553",
   publisher = "Annual Reviews",
   issn = "1545-4118",
   type = "Journal Article",
  }

@article{rajca1994organic,
  author={Rajca, Andrzej},
title={Organic Diradicals and Polyradicals: From Spin Coupling to Magnetism?},
journal={Chemical Reviews},
year={1994},
month={Jun},
day={01},
publisher={American Chemical Society},
volume={94},
number={4},
pages={871-893},
issn={0009-2665},
doi={10.1021/cr00028a002},
url={https://doi.org/10.1021/cr00028a002}
}

@Article{Shirai2023,
author={Shirai, Soichi
and Iwakiri, Hokuto
and Kanno, Keita
and Horiba, Takahiro
and Omiya, Keita
and Hirai, Hirotoshi
and Koh, Sho},
title={Computational Analysis of Chemical Reactions Using a Variational Quantum Eigensolver Algorithm without Specifying Spin Multiplicity},
journal={ACS Omega},
year={2023},
month={Jun},
day={06},
publisher={American Chemical Society},
volume={8},
number={22},
pages={19917-19925},
doi={10.1021/acsomega.3c01875},
url={https://doi.org/10.1021/acsomega.3c01875}
}

@article{kattem2022variational,
  title = {Variational quantum eigensolver for the Heisenberg antiferromagnet on the kagome lattice},
  author = {Kattem\"olle, Joris and van Wezel, Jasper},
  journal = {Phys. Rev. B},
  volume = {106},
  issue = {21},
  pages = {214429},
  numpages = {17},
  year = {2022},
  month = {Dec},
  publisher = {American Physical Society},
  doi = {10.1103/PhysRevB.106.214429},
  url = {https://link.aps.org/doi/10.1103/PhysRevB.106.214429}
}

@Article{Yi2025,
author={Yi, Xin
and Huo, Jiacheng
and Liu, Guanhua
and Fan, Ling
and Zhang, Ru
and Cao, Cong},
title={A probabilistic quantum algorithm for imaginary-time evolution based on Taylor expansion},
journal={EPJ Quantum Technology},
year={2025},
month={Apr},
day={03},
volume={12},
number={1},
pages={43},
issn={2196-0763},
doi={10.1140/epjqt/s40507-025-00347-0},
url={https://doi.org/10.1140/epjqt/s40507-025-00347-0}
}

@misc{qiskit2024,
      title={Quantum computing with {Q}iskit},
      author={Javadi-Abhari, Ali and Treinish, Matthew and Krsulich, Kevin and Wood, Christopher J. and Lishman, Jake and Gacon, Julien and Martiel, Simon and Nation, Paul D. and Bishop, Lev S. and Cross, Andrew W. and Johnson, Blake R. and Gambetta, Jay M.},
      year={2024},
      doi={10.48550/arXiv.2405.08810},
      eprint={2405.08810},
      archivePrefix={arXiv},
      primaryClass={quant-ph}
}

@article{Mootz2024,
doi = {10.1088/2058-9565/ad57ea},
url = {https://dx.doi.org/10.1088/2058-9565/ad57ea},
year = {2024},
month = {jun},
publisher = {IOP Publishing},
volume = {9},
number = {3},
pages = {035054},
author = {Mootz, Martin and Orth, Peter P and Huang, Chuankun and Luo, Liang and Wang, Jigang and Yao, Yong-Xin},
title = {Two-dimensional coherent spectrum of high-spin models via a quantum computing approach},
journal = {Quantum Science and Technology},
}

@inproceedings{gilyen2019quantum,
author = {Gily\'{e}n, Andr\'{a}s and Su, Yuan and Low, Guang Hao and Wiebe, Nathan},
title = {Quantum singular value transformation and beyond: exponential improvements for quantum matrix arithmetics},
year = {2019},
isbn = {9781450367059},
publisher = {Association for Computing Machinery},
address = {New York, NY, USA},
url = {https://doi.org/10.1145/3313276.3316366},
doi = {10.1145/3313276.3316366},
booktitle = {Proceedings of the 51st Annual ACM SIGACT Symposium on Theory of Computing},
pages = {193–204},
numpages = {12},
location = {Phoenix, AZ, USA},
series = {STOC 2019}
}

@article{albash2018adiabatic,
  title = {Adiabatic quantum computation},
  author = {Albash, Tameem and Lidar, Daniel A.},
  journal = {Rev. Mod. Phys.},
  volume = {90},
  issue = {1},
  pages = {015002},
  numpages = {64},
  year = {2018},
  month = {Jan},
  publisher = {American Physical Society},
  doi = {10.1103/RevModPhys.90.015002},
  url = {https://link.aps.org/doi/10.1103/RevModPhys.90.015002}
}

@Article{vasiliev2018milestpnes,
author={Vasiliev, Alexander
and Volkova, Olga
and Zvereva, Elena
and Markina, Maria},
title={Milestones of low-D quantum magnetism},
journal={npj Quantum Materials},
year={2018},
month={Mar},
day={28},
volume={3},
number={1},
pages={18},
issn={2397-4648},
doi={10.1038/s41535-018-0090-7},
url={https://doi.org/10.1038/s41535-018-0090-7}
}

@book{dalzell2025quantum, 
place={Cambridge}, 
title={Quantum Algorithms: A Survey of Applications and End-to-end Complexities}, 
publisher={Cambridge University Press}, 
author={Dalzell, Alexander M. and McArdle, Sam and Berta, Mario and Bienias, Przemyslaw and Chen, Chi-Fang and Gilyén, András and Hann, Connor T. and Kastoryano, Michael J. and Khabiboulline, Emil T. and Kubica, Aleksander and et al.}, year={2025},
doi={10.1017/9781009639651}
}

@article{barenco1995elementary,
  title = {Elementary gates for quantum computation},
  author = {Barenco, Adriano and Bennett, Charles H. and Cleve, Richard and DiVincenzo, David P. and Margolus, Norman and Shor, Peter and Sleator, Tycho and Smolin, John A. and Weinfurter, Harald},
  journal = {Phys. Rev. A},
  volume = {52},
  issue = {5},
  pages = {3457--3467},
  numpages = {0},
  year = {1995},
  month = {Nov},
  publisher = {American Physical Society},
  doi = {10.1103/PhysRevA.52.3457},
  url = {https://link.aps.org/doi/10.1103/PhysRevA.52.3457}
}

@ARTICLE{william2025halving,
       author = {{Simon}, William A. and {Love}, Peter J.},
        title = "{Halving the Cost of Controlled Time-Evolution}",
      journal = {arXiv e-prints},
         year = 2025,
        month = nov,
          eid = {arXiv:2511.13855},
        pages = {arXiv:2511.13855},
          doi = {10.48550/arXiv.2511.13855},
archivePrefix = {arXiv},
       eprint = {2511.13855},
 primaryClass = {quant-ph},
       adsurl = {https://ui.adsabs.harvard.edu/abs/2025arXiv251113855S}
}

@article{dong2022ground,
  title = {Ground-State Preparation and Energy Estimation on Early Fault-Tolerant Quantum Computers via Quantum Eigenvalue Transformation of Unitary Matrices},
  author = {Dong, Yulong and Lin, Lin and Tong, Yu},
  journal = {PRX Quantum},
  volume = {3},
  issue = {4},
  pages = {040305},
  numpages = {25},
  year = {2022},
  month = {Oct},
  publisher = {American Physical Society},
  doi = {10.1103/PRXQuantum.3.040305},
  url = {https://link.aps.org/doi/10.1103/PRXQuantum.3.040305}
}

@article{shijie2020full,
author = {Shijie Wei  and Hang Li  and GuiLu Long },
title = {A Full Quantum Eigensolver for Quantum Chemistry Simulations},
journal = {Research},
volume = {2020},
number = {},
pages = {},
year = {2020},
doi = {10.34133/2020/1486935},
URL = {https://spj.science.org/doi/abs/10.34133/2020/1486935},
eprint = {https://spj.science.org/doi/pdf/10.34133/2020/1486935},
}

@article{wang2024improving,
  title = {Improving the full quantum eigensolver with exponentiated operators},
  author = {Wang, Bozhi and Wen, Jingwei and Wu, Jiawei and Xie, Haonan and Yang, Fan and Ruan, Dong and Wei, Shijie and Long, Gui-lu},
  journal = {Phys. Rev. B},
  volume = {109},
  issue = {24},
  pages = {245117},
  numpages = {17},
  year = {2024},
  month = {Jun},
  publisher = {American Physical Society},
  doi = {10.1103/PhysRevB.109.245117},
  url = {https://link.aps.org/doi/10.1103/PhysRevB.109.245117}
}
\onecolumngrid
\appendix
\section{Binary encoding of $S=5/2$ and $S=2$ Spin states}\label{sec:binary}
Here, we present the explanation of the binary encoding for $S=5/2$ and $S=2$, which is used in numerical experiments of the manganese trimer. There are several methods to encode high-spin states, such as Gray encoding, but in this paper, we used the standard binary encoding. In the standard binary encoding, the transformation from original states to encoded states is summarized in Tab.~\ref{tab:binary_encoding}.
\begin{table}[h]
\centering
\begin{tabular}{|c|c||c|c|}
\hline
\textbf{Spin state} & \textbf{Encoded} & \textbf{Spin state} & \textbf{Encoded} \\ \hline
$\ket{5/2, 5/2}$      & $\ket{000}$      & $\ket{2, 2}$      & $\ket{000}$      \\ \hline
$\ket{5/2, 3/2}$      & $\ket{001}$      & $\ket{2, 1}$      & $\ket{001}$      \\ \hline
$\ket{5/2, 1/2}$      & $\ket{010}$      & $\ket{2, 0}$      & $\ket{010}$      \\ \hline
$\ket{5/2, -1/2}$      & $\ket{011}$      & $\ket{2, -1}$      & $\ket{011}$      \\ \hline
$\ket{5/2, -3/2}$      & $\ket{100}$      & $\ket{2, -2}$      & $\ket{100}$      \\ \hline
$\ket{5/2, -5/2}$      & $\ket{101}$      &       &       \\ \hline
\end{tabular}
\caption{Binary encoding representation of $S=5/2$ and $S=2$ Spin states.}
\label{tab:binary_encoding}
\end{table}

The spin operators are also rewritten. The encoded spin operators for $S=5/2$ can be written as follows.
In this appendix, for the sake of clarity in notation, Pauli matrices $\sigma_{i x}$, $\sigma_{i y}$, and $\sigma_{i z}$ are written as $X_{i}$, $Y_{i}$, and $Z_{i}$, respectively.
\begin{align*}
    S_{x} &= \frac{3+2\sqrt{5}}{8}X_{0}+\frac{-3+2\sqrt{5}}{8}Z_{1}X_{0}+\frac{3}{8}(Z_{2}X_{0}-Z_{2}Z_{1}X_{0})\\
    &+\frac{1}{\sqrt{8}}(X_{1}X_{0}+Y_{1}Y_{0}+X_{2}X_{1}X_{0}-X_{2}Y_{1}Y_{0}+Y_{2}X_{1}Y_{0}+Y_{2}Y_{1}X_{0}+Z_{2}X_{1}X_{0}+Z_{2}Y_{1}Y_{0})\stepcounter{equation}\tag{\theequation} \\
    S_{y} &= \frac{3+2\sqrt{5}}{8}Y_{0}+\frac{-3+2\sqrt{5}}{8}Z_{1}Y_{0}+\frac{3}{8}(Z_{2}Y_{0}-Z_{2}Z_{1}Y_{0})\\
    &+\frac{1}{\sqrt{8}}(Y_{1}X_{0}-X_{1}Y_{0}-Y_{2}Y_{1}Y_{0}+Y_{2}X_{1}X_{0}-X_{2}Y_{1}X_{0}-X_{2}X_{1}Y_{0}+Z_{2}Y_{1}X_{0}-Z_{2}X_{1}Y_{0})\stepcounter{equation}\tag{\theequation} \\
    S_{z} &= \frac{1}{8}(Z_{2}Z_{0}+Z_{1}Z_{0}-Z_{2}Z_{1}Z_{0})+\frac{3}{8}Z_{0}+(Z_{2}+Z_{2}Z_{1})\stepcounter{equation}\tag{\theequation} \\
\end{align*}
The encoded spin operators for $S=2$ can be written like
\begin{align*}
    S_{x} &= \frac{1+\sqrt{3/2}}{4}(X_{0}+Z_{2}X_{0})+\frac{1-\sqrt{3/2}}{4}(Z_{1}X_{0}+Z_{2}Z_{1}X_{0}) +\frac{\sqrt{3/2}}{4}(X_{1}X_{0}+Y_{1}Y_{0}+Z_{2}X_{1}X_{0}+Z_{2}Y_{1}Y_{0})\\
    &+ \frac{1}{4}(X_{2}X_{1}X_{0}-X_{2}Y_{1}Y_{0}+Y_{2}X_{1}Y_{0}+Y_{2}Y_{1}X_{0})
    \stepcounter{equation}\tag{\theequation}\\ 
    S_{y} &= \frac{1+\sqrt{3/2}}{4}(Y_{0}+Z_{2}Y_{0})+\frac{1-\sqrt{3/2}}{4}(Z_{1}Y_{0}+Z_{2}Z_{1}Y_{0}) +\frac{\sqrt{3/2}}{4}(Y_{1}X_{0}-X_{1}Y_{0}+Z_{2}Y_{1}X_{0}-Z_{2}X_{1}Y_{0})\\
    &+ \frac{1}{4}(Y_{2}X_{1}X_{0}-X_{2}X_{1}Y_{0}-X_{2}Y_{1}X_{0}-Y_{2}Y_{1}Y_{0})\stepcounter{equation}\tag{\theequation} \\
    S_{z} &= \frac{1}{4}(Z_{1}-Z_{1}Z_{0}+Z_{2}Z_{1}Z_{0}) +\frac{1}{2}(Z_{2}+Z_{2}Z_{0})+\frac{3}{4}Z_{2}Z_{1}
    \stepcounter{equation}\tag{\theequation} 
\end{align*}
Note that in the numerical experiments, the evaluation does not reveal any differences in the results due to the choice of encoding method, even though the encoding affects the number of qubits, gate complexity, and robustness to qubit noise.
\end{document}